\documentclass[12pt]{article}

\usepackage{amsmath}
\usepackage{amssymb}
\usepackage{amsfonts}
\usepackage{graphicx}

\numberwithin{equation}{section}

\begin{document}

\begin{center}
{\large \bf{ The Higgs condensate as a quantum liquid: A critical comparison with observations}}
\end{center}

\vspace*{2cm}

\begin{center}
{
Paolo Cea~\protect\footnote{Electronic address:
{\tt paolo.cea@ba.infn.it}}  \\[0.5cm]
{\em INFN - Sezione di Bari, Via Amendola 173 - 70126 Bari,
Italy} }
\end{center}

\vspace*{1.5cm}

\begin{abstract}
\noindent 
The triviality of four-dimensional scalar quantum field theories poses challenging problems to the
usually adopted perturbative implementation of the Higgs mechanism. In the first part of the paper
we compare the triviality  scenario and the renormalised two-loop perturbation theory to precise and extensive results
from non-perturbative numerical simulations of the real scalar field theory on the lattice. The proposal of triviality
and spontaneous symmetry breaking turns out to be in good agreement with numerical simulations, while
the renormalised perturbative approach seems to suffer  significant deviations from the numerical 
simulation results. In the second part of the paper  we try to illustrate how the
triviality of four-dimensional scalar field theory leads,  nevertheless, to the spontaneous symmetry breaking
in the scalar sector of the Standard Model. We show how  triviality  allows us to develop a physical picture
of the Higgs mechanism in the Standard Model. We suggest that the Higgs condensate behave like a relativistic
quantum liquid leading to the prevision of  two Higgs bosons. The light  Higgs boson   resembles closely
the new LHC narrow resonance at 125 GeV. The heavy Higgs boson is a rather broad resonance with
mass of about 730 GeV. We critically compare our proposal to the complete LHC Run 2 data collected
by the ATLAS and CMS Collaborations.  We do not find convincingly evidences of the heavy Higgs boson in
the ATLAS datasets. On the other hand, the CMS full Run 2 data  display evidences of a heavy Higgs boson in
the main decay modes $H \rightarrow WW$, while in the preliminary Run 2 data there are hints  of the decays  $H \rightarrow ZZ$ 
 in the golden channel. We also critically discuss plausible reasons for the discrepancies between the two LHC experiments.
\vspace{0.5cm}

\vspace{0.2cm}


\end{abstract}
\newpage

\noindent
\section{Introduction}
\label{s-1}
 Scalar quantum field theories  in four dimensions are of fundamental  interest  since they play a fundamental role in the Standard Model 
 giving rise to the   generation of fermion and gauge boson masses via the Higgs 
 mechanism~\cite{Englert:1964,Higgs:1964,Guralnik:1964,Higgs:1966}.  
In 2012 a new particle was discovered by the ATLAS and CMS experiments at the Large Hadron Collider (LHC)~\cite{Aad:2012,Chatrchyan:2012}.
This new particle, with a mass of about 125 GeV, turned out to be consistent with the Standard Model 
Higgs boson thereby validating  the Higgs mechanism.  \\ 
In the Standard Model of Particle Physics the Higgs mechanism is implemented  by introducing  scalar fields, the Higgs fields, 
with a specific potential.  In the generally accepted  picture,  the spontaneous symmetry breaking in the Standard Model is 
managed  within the perturbation theory which leads to the prediction  that the Higgs boson mass squared  is proportional to $\lambda \, v^2$,
where $\lambda$ is the positive scalar self-coupling and  $v \simeq 246$  GeV  is the known weak scale. 
To do this, first one assume that the quantum Higgs fields undergo the condensation by means of a tachyonic mass term $\mu^2 < 0$.
After that, it is assumed the existence of a short-range repulsion, assured by the positive quartic self-coupling, to stabilise
the condensation of the Higgs fields. This description of a fundamental aspect of the Standard Model is highly  unsatisfactory.
The first problem comes out from the tachyonic mass term. Indeed, there are no physical mechanisms able to generate a negative squared mass.
So that assuming the presence of the tachyonic mass is simply unjustified. In other words, one is assuming that for some reasons 
the Higgs fields are condensing in the ground state. The second  problem arises from the fact that  self-interacting scalar fields in four space-time
dimensions  are subject to the triviality problem~\cite{Fernandez:1992}, namely the renormalised self-coupling
$\lambda \rightarrow  0$ when the ultraviolet cutoff  is sent to infinity.    Strictly speaking, up to now there was  no rigorous proof 
of triviality.  However,   extensive  non-perturbative numerical studies  in four dimensions  had convincingly confirmed  the triviality conjecture.
On the other hand, it is worthwhile to mention that  in the recent paper Ref.~\cite{Aizenman:2021} it has been rigorously demonstrated
the triviality of the scaling limits of the Ising and self-interacting scalar field models in four dimensions.
Namely,  the scaling limits of  the critical Ising and self-interacting scalar field in four dimensions are Gaussian
as Euclidean field theories. Thus, we see that there are no more doubts on the Gaussian, or triviality,  nature of one-component
self-interacting scalar quantum field theory. \\
Actually, in the usual implementation of the renormalised perturbation theory it is  assumed that trivial scalar field theories exist and may be far from
being free provided one  allows  for a large but finite ultraviolet cutoff $\Lambda$. For instance, the cutoff $\Lambda$ may be introduced explicitly
 by a lattice with finite lattice spacing $a$. In any case, in Minkowski  space-time one must admit that the momenta are constrained by:
\begin{equation} 
\label{1.1} 
\left | \; k_{\mu} \;  \right |  \; \lesssim  \; \Lambda  \; \; \; , \; \; \; \mu \; = \;  0, 1, 2, 3   \; ,
\end{equation}
 while, equivalently, in  the Euclidean field theory on a hypercubic lattice:
\begin{equation} 
\label{1.2} 
\left | \; k_{i} \;  \right |  \; \lesssim  \; \Lambda \;  \simeq \; \frac{\pi}{a}   \; \; \; , \; \; \; i \; = \;  1, 2, 3, 4    \; .
\end{equation}
The innocently looking condition Eq.~(\ref{1.1}), or the Euclidean version Eq.~(\ref{1.2}), is routinely employed  in 
perturbative calculations by means of Feynman diagrams. However, quantum field theories are not a mere collection of
Feynman diagrams.  The general theory of relativistic quantised fields is formulated in terms of fields $\hat{\phi}(x)$
that are operator-valued tempered distributions densely defined in the Hilbert space of physical states (see, for instance, 
Refs.~\cite{Streater:1964,Strocchi:1993}). The singular behaviour of the fields as a function of the spacetime point $x$ is an
inevitable consequence of relativistic covariance. In fact, there is a pertinent theorem due to Wightman~\cite{Wightman:1964}
that says that, if one assumes that a quantum field exists as an (in general unbounded) operator at a spacetime point $x$ and it is covariant
with respect to a (strongly continuous) unitary representation of the Poincare group, then the quantum field is trivial in the sense
that it is a multiple of the identity. In other words, the Hilbert space of the physical states should consist only of the vacuum state.
It is worthwhile to stress that the theorem holds only if the quantum field is local. Thus, we may evade the Wightman theorem if 
the quantum field is non-elementary or compound. \\
By the Wightman theorem we known that the local quantum field must be singular. More precisely, we already said that the
quantum field operator $\hat{\phi}(x)$ must be a (tempered) distribution whose singularities are at most finite
derivatives of the Dirac $\delta$ distribution.  If we admit the validity of Eq.~(\ref{1.1}), or Eq.~(\ref{1.2}), then the Dirac
$\delta$ distribution becomes a non-singular fat $\delta$-function and, therefore, the quantum scalar field is a non-singular operator
so that the Hilbert space of the physical states becomes empty. This shows that we cannot evade the triviality problem for relativistic
local scalar quantum fields. The unique way out left would be to consider the Higgs scalar fields as non-elementary.
In this case the cutoff $\Lambda$ must be interpreted as a high-energy scale such that, for distances smaller than $1/\Lambda$,
the Higgs fields are replaced by unknown elementary quantum fields. However, the problem is that, up to now, the experimental observations at the Large Hadron Collider did not displayed any signs of physics beyond the Standard Model. So that we are led to exclude these
mundane possibilities. Nevertheless, it must be mentioned that recently~\cite{Romatschke:2023} it has been suggested that 
 multi-component scalar field theories in four dimensions  may evade  triviality  if the quartic self-coupling constant  could  approach
 zero from below in the continuum limit. \\
There is another problematic aspect that affects the perturbative treatment of the Higgs mechanism. As a matter of fact,
the naive perturbation theory overlooks the fact that, according to the Haag's theorem~\cite{Haag:1955}~\footnote{ The interested readers 
can found a rather complete account on the Haag's theorem in Refs.~\cite{Earman:2006,Klaczynski:2016,Duncan:2012,Strocchi:2013}},
the quantum vacuum with scalar condensate is inequivalent to the perturbative quantum vacuum, i.e. the free massive scalar field
quantum vacuum. Here inequivalent means that the two vacua are nor connected by a unitary transformation.
In  particular, as we shall discuss later on, the presence of the scalar condensate imposes a constraint on the scalar quantum field
that, according to a seminal paper by K. Symanzik~\cite{Symanzik:1981} on the Schr\"odinger representation and Casimir effect in quantum field theories, requires a renormalisation of the condensate strength that is different from the usual wavefunction renormalisation
constant. This should completely  jeopardise  the perturbative approach such that the renormalised perturbative calculations could lead to misleading results. \\
 The aim of the present paper is twofold.  Firstly, in Sect.~\ref{s-2} we argue that spontaneous symmetry breaking in a  relativistic 
local quantum scalar field theory is not incompatible with the triviality of the theory. To this end, we focus on a real massless scalar
 field and show that quantum fluctuations can drive the scalar condensation if one admits an even infinitesimal positive quartic
self-coupling. The essence of the mechanism practically coincides with the one-loop evaluation of the effective potential performed
in the classical paper by Coleman and Weinberg~\cite{Coleman:1973} (see, also, Ref.~\cite{Weinberg:1973}). 
Note that the suggestion of the coexistence of spontaneous symmetry
 breaking with triviality has been already advanced since long time in Ref.~\cite{Consoli:1994,Consoli:1997}.
In Sect.~\ref{s-3} we compare  the triviality and symmetry breaking scenario as well as the renormalised two-loop perturbation theory approach
to precise and extensive results from non-perturbative numerical simulations of the real scalar field theory on the lattice. Here we shall see
that the triviality and spontaneous symmetry breaking proposal is in quite good agreements with the available results from numerical simulations,
while the renormalised perturbative approach displays significant deviations from the numerical simulation outcomes. \\
In the second part of the paper, we apply the triviality and spontaneous symmetry breaking mechanism to the scalar sector of the Standard
Model. We develop a physical picture of the Higgs mechanism in the Standard Model which is not subject to
the naturalness and stability problems.  In Sect.~\ref{s-4} we further illustrate the suggestion, already advanced in a previous
 paper~\cite{Cea:2020}, that the Higgs condensate behaves as the relativistic version of a quantum liquid much like the He II
 in the  superfluid phase.    This proposal lead to the remarkable result that the Higgs condensate excitations behave as two Higgs bosons. 
 The light Higgs boson was already   identified with the new LHC narrow resonance at 125 GeV~\cite{Cea:2020}. 
 On the other hand, the heavy Higgs boson   turns out to a broad resonance with central mass of about 730 GeV. 
  In Sect.~\ref{s-5} we critically contrast our theoretical proposal    to the complete Run 2 datasets from both ATLAS and CMS experiments. 
  In particular, we looked at evidences for our heavy Higgs boson.
 We anticipate that we did not find convincingly evidences of the heavy Higgs boson in the ATLAS data (Sect.~\ref{s-5-1}). 
 On the contrary, the CMS data (Sect.~\ref{s-5-2}) seem to display reasonable evidences of a heavy Higgs boson in the main decay channels 
 $H \rightarrow VV$, V=W, Z  that  compare quite favourably to our theoretical proposal. Finally, in Sect.~\ref{s-6} we briefly summarise the main results of the paper and  draw our conclusions. 
\section{Triviality and spontaneous symmetry breaking}
\label{s-2}
In this section we would like to illustrate the triviality and spontaneous symmetry breaking scenario within the simplest scalar
quantum field theory, namely the one-component quantum scalar field $\hat{\phi}(x)$. To start with, we  recall  the
spontaneous symmetry breaking mechanism  in the standard  treatment. Let us consider the Lagrangian: 
\begin{equation} 
\label{2.1} 
{\cal{L}}\;  = \; \frac{1}{2} \, (\partial_\mu \hat{\phi})^2 \; - \; \frac{1}{2} \, \mu^2  \,  \hat{\phi}^2 \; - \;
 \frac{1}{4} \,  \lambda_0  \,  \hat{\phi}^4  \; ,
\end{equation}
where $\lambda_0$ and $\hat{\phi}$ are the bare quartic self-coupling and quantum field respectively. The mass parameter $\mu^2$ is 
assumed to be negative (tachyonic mass term) so that the quantum scalar field is forced to develops a non-zero vacuum expectation
value $\phi$. Due to the translational invariance of the quantum vacuum $\phi$ must be a real number, i.e. the scalar condensate is
time-independent and spatially uniform. After that  the true quantum vacuum is obtained by writing  $\phi = v$, where $v$ is obtained by
minimising the classical potential term in the Lagrangian Eq.~(\ref{2.1}), and 
\begin{equation} 
\label{2.2} 
\hat{\phi}(x)  \; =  \;  \hat{\eta}(x)  \; +  \;  v \; .
\end{equation}
One readily obtains that the shifted quantum field  $\hat{\eta}(x)$ is a massive scalar field with mass  $m^2 = 2  \lambda_0 \, v^2$
and with quartic and cubic self-interaction terms. 
However, as we said in the Introduction, this approach is highly questionable since  there are no physical mechanisms
 able to generate a negative squared mass for a local quantum scalar field. We are led, therefore, to assume $\mu^2 = 0$. In this case
 the minimum of the classical potential is at $v = 0$. The only way out left is to hope that quantum fluctuations could induce the vacuum
 condensation of the scalar field. To check  this we need to evaluate the effective potential that includes the quantum corrections to the classical
 potential. The effective potential can be systematically evaluated in an expansion in power of the Planck's constant.  The lowest order approximation  corresponds to the one-loop approximation.
As it is well known~\cite{Coleman:1973}, the lowest order  effective potential is obtained by summing the one-loop vacuum diagrams:
\begin{equation} 
\label{2.3} 
V_{1-loop}(\phi)  =  \frac{1}{4} \lambda_0 \, \phi^4  - \frac{i}{2}  \int \frac{d^4 k}{(2 \pi)^4} \;  \ln  [ - k_0^2 + \vec{k}^2  + 3 \lambda_0 \,  \phi^2  
- i \epsilon ] \, .
\end{equation}
Integrating over $k_0$ and discarding an inessential  constant gives:
\begin{equation} 
\label{2.4} 
V_{1-loop}(\phi)  =  \frac{1}{4} \lambda_0 \, \phi^4  + \frac{1}{2}  \int \frac{d^3 k}{(2 \pi)^3}  \;  \sqrt{  \vec{k}^2  + 3 \lambda_0 \,   \phi^2  } \, .
\end{equation}
This last equation offers an interesting interpretation of the one-loop effective potential.  In fact, Eq.~(\ref{2.4}) shows that the 
$V_{1-loop}(\phi)$ can be interpreted as the free  vacuum energy density of the shifted field $ \hat{\eta}(x)$. To see this  we note that
in the quadratic approximation the Hamiltonian of the fluctuation  over the condensate background $\hat{\eta}$  is given by:
\begin{equation} 
\label{2.5} 
\hat{\cal{H}}_0  = \frac{1}{2}  \hat{\Pi}_{\eta}^2(x) \; + \; \frac{1}{2}  (\vec{\nabla} \hat{\eta}(x))^2 \; + \; \frac{1}{2}  \; ( 3 \lambda_0 \,  \phi^2 ) \;
\hat{ \eta}^2(x) \;  + \; \frac{1}{4} \lambda_0 \, \phi^4 \; ,
\end{equation}
where $ \hat{\Pi}_{\eta}(x)$ is the  conjugate momentum of the fluctuating field. \\
Performing the integrals in Eq.~(\ref{2.4}) and introducing an ultraviolet cutoff $\Lambda$, after mass
 renormalisation~\cite{Coleman:1973,Weinberg:1973}, one obtains easily:
\begin{equation} 
\label{2.6} 
V_{1-loop}(\phi)  =  \frac{1}{4} \lambda_0 \, \phi^4  + \frac{\omega(\phi)^4}{64 \pi^2}  \ln \left ( \frac{ \omega^2(\phi)}{\Lambda^2} \right )   \;  \;  , \; \;  
\omega^2(\phi) \, = \,   3 \lambda_0 \,  \phi^2 \; .
\end{equation}
A straightforward calculation shows that  the one-loop effective potential displays a minimum at:
\begin{equation} 
\label{2.7} 
 3 \lambda_0 \,  v^2 \; = \;   \frac{\Lambda^2}{\sqrt{e}} \; \exp{ \left [-   \frac{16 \pi^2}{9 \lambda_0} \right ]} \; .
\end{equation}
Indeed, one can easily check that:
\begin{equation} 
\label{2.8} 
V_{1-loop}(v)   \; = \;  - \;  \frac{\omega^4(v)}{128 \pi^2 } \;  \; ,
\end{equation} 
Using Eq.~(\ref{2.7}) we can rewrite Eq.~(\ref{2.6}) as:
\begin{equation} 
\label{2.9} 
V_{1-loop}(\phi)  \; = \;  \frac{\omega^4(\phi)}{64 \pi^2} \left [  \ln \left ( \frac{ \phi^2}{v^2} \right ) - \frac{1}{2} \right ]  \; .
\end{equation}
Since we are dealing with a renormalisable theory,  the effective potential   must satisfy the  renormalisation group invariance. So
that  for $\Lambda \rightarrow \infty$  we have:
\begin{equation} 
\label{2.10} 
\left [   \Lambda \frac{\partial }{\partial  \Lambda} \; + \; \beta \, \frac{\partial }{\partial  \lambda_0}   \; + \; \gamma \, \phi \, 
 \frac{\partial }{\partial  \phi}
          \right ] V_{1-loop}(\phi)  \; = \; 0   \; ,
\end{equation}
where $\gamma$ is the anomalous dimension of the quantum scalar field  related to the wavefunction renormalisation constant and
the beta-function is defined by:
\begin{equation} 
\label{2.11} 
  \Lambda \frac{\partial }{\partial  \Lambda} \;  \lambda  \; = \; \beta(\lambda)   \; .
\end{equation}
In the previous equations we are considering the bare coupling constant as the quartic self-coupling at the cutoff scale $\Lambda$.
The renormalisation process tells us that we can change the bare coupling $\lambda_0$ and the  ultraviolet  cutoff 
$\Lambda$  while, at the same time, keeping the renormalised coupling constant fixed at some value.
Within perturbation theory one finds~\cite{Coleman:1973,Weinberg:1973}:
\begin{equation} 
\label{2.12} 
\gamma_{pert}  \; = \; 0 \; \; \; \; , \; \; \; \; \beta_{pert}(\lambda) \;= \; \frac{9}{8 \pi^2} \,  \lambda^2 \; .
\end{equation}
Thus, the one-loop corrections have generated spontaneous symmetry breaking. However, the minimum of the effective potential lies outside the expected range of validity of the one-loop approximation and it must be rejected as an artefact of the approximation.  Indeed, the renormalisation
group resummed effective potential does not display symmetry breaking. 
On the other hand, as discussed in  Section~\ref{s-1}, there are no doubt on the triviality of the theory. As a consequence, within perturbation theory there is no room for symmetry breaking in a self-interacting scalar field theory. 
However, following the suggestion of Refs.~\cite{Consoli:1994,Consoli:1997} we argue below that spontaneous symmetry breaking
could be compatible  with the triviality of the scalar quantum field theory.  
 The arguments go as follows. Let us suppose that, notwithstanding  the triviality problem, the massless quantum scalar field $\hat{\phi}(x)$
 undergoes the vacuum condensation.  This means that, according to Eq.~(\ref{2.2}), the quantum field can be decomposed into the
 fluctuating  quantum field  $\hat{\eta}(x)$ and the uniform vacuum condensate $\phi$. In this case the triviality of the theory means
 that the quantum fluctuation must be a free field. The presence of the vacuum scalar condensate suggests that  $\hat{\eta}(x)$
is a massive free field.  Our previous one-loop  perturbative discussion indicated that the vacuum condensation and scalar
mass generation of the fluctuating field could originate from vacuum quantum fluctuations.  Now, the quantum fluctuations of
free field are Gaussian and, thereby, cannot induce dynamically the spontaneous symmetry breaking we are looking for.
Our previous discussion of the Coleman and Weinberg mechanism showed that the dynamical spontaneous symmetry breaking
requires the presence of non-Gaussian vacuum fluctuations introduced by a positive quartic self-coupling $\lambda_0 > 0$.
The presence of the quartic self-coupling  leads to the vacuum condensation of the scalar field and gives  rise to a mass term for 
the free fluctuating quantum field without  introducing  any new interaction terms  to be consistent  with the triviality nature
of the quantum scalar field.  Obviously, the requirement of  a finite quartic self-coupling is in  striking contradiction with
the triviality of the theory. For the time being, we simply adopt we above assumptions. It should be clear that, to be consistent,
we must send the quartic self-coupling to zero at the end of our calculations. We will return on this delicate point later on. \\
From the above discussion it follows that the quantum Hamiltonian of the quantum field $\hat{\eta}(x)$ is the free
Hamiltonian operator:
\begin{equation} 
\label{2.13} 
\hat{\cal{H}}  = \frac{1}{2}  \hat{\Pi}_{\eta}^2(x) \; + \; \frac{1}{2}  (\vec{\nabla} \hat{\eta}(x))^2 \; + \; \frac{1}{2}  \; \tilde{\omega}^2(\phi) \;
\hat{ \eta}^2(x) \;  + \; \frac{1}{4} \lambda_0 \, \phi^4 \; ,
\end{equation}
where:
\begin{equation} 
\label{2.14} 
\tilde{\omega}^2(\phi) \; = \;  3 \, \tilde{\lambda} \,  \phi^2  \;   \; , \; \;  \tilde{\lambda} \; = \; a_1 \, \lambda_0 \; \; ,
\end{equation}
where $a_1$ is some numerical constant. Note that for $a_1 =1$ we recover the one-loop effective Hamiltonian Eq.~(\ref{2.5}),
while $a_1 = 2/3$ corresponds to the variational Gaussian approximation. The actual value of the parameter $a_1$ will be
fixed in the next Section by comparing  with non-perturbative numerical simulations of the Euclidean version of the quantum
scalar theory.
Since the quantum Hamiltonian is quadratic in the fields,  the {\it{ exact}} effective potential is simply the vacuum energy density:
\begin{equation} 
\label{2.15} 
V_{triv}(\phi)  =  \frac{1}{4} \lambda_0 \, \phi^4  + \frac{1}{2}  \int \frac{d^3 k}{(2 \pi)^3}  \;  \sqrt{  \vec{k}^2  + \tilde{\omega}^2(\phi)  } \; = \;
 \frac{1}{4} \lambda_0 \, \phi^4  + \frac{\tilde{\omega}^4(\phi)}{64 \pi^2}  \ln \left ( \frac{ \tilde{\omega}^2(\phi)}{\Lambda^2} \right )   \;  \; .
\end{equation}
 As in the  previous one-loop calculations  the triviality  effective potential displays a minimum at:
\begin{equation} 
\label{2.16} 
 3 \tilde{\lambda}  v^2 \; = \;   \frac{\Lambda^2}{\sqrt{e}} \; \exp{ [-   \frac{16 \pi^2}{9 \tilde{\lambda}}]} \; .
\end{equation}
Introducing:
\begin{equation} 
\label{2.17} 
 m_H^2 \; = \; \tilde{\omega}^2(v) \; = \;   3 \tilde{\lambda}  v^2   \; ,
\end{equation} 
we recast the effective potential into the form:
\begin{equation} 
\label{2.18} 
V_{eff}(\phi)  =  \frac{(3  \tilde{\lambda} \,  \phi^2)^2}{64 \pi^2}    \; \left [
  \ln \left ( \frac{ 3  \tilde{\lambda} \,  \phi^2}{m_H^2} \right ) \;  - \; \frac{1}{2}  \right ] \;  \; .
\end{equation}
This last equation shows that:
\begin{equation} 
\label{2.19} 
V_{triv}(v)   \; = \;  - \;  \frac{m_H^4}{128 \pi^2 } \;  \; , 
\end{equation} 
confirming that the effective potential has a negative minimum at $\phi = v$. 
Now the problem is to see if it exists the continuum limit $\Lambda \rightarrow \infty$  and if this limit is consistent with
the triviality of the theory. Obviously, we must have:
\begin{equation} 
\label{2.20} 
\left [   \Lambda \frac{\partial }{\partial  \Lambda} \; + \; \beta(\lambda_0) \, \frac{\partial }{\partial  \lambda_0}   \; + \; \gamma(\lambda_0) \, \phi \,  \frac{\partial }{\partial  \phi}
          \right ] V_{triv}(\phi)  \; = \; 0   \; .
\end{equation}
Note that, at variance of the perturbative approach,  we cannot use perturbation theory to determine 
$\beta(\lambda_0)$ and $\gamma(\lambda_0)$.  First, we  may use 
 Eq.~(\ref{2.20}) at the minimum $v$  where the derivative of the effective potential vanishes. We get:
\begin{equation} 
\label{2.21} 
\left [   \Lambda \frac{\partial }{\partial  \Lambda} \; + \; \beta(\lambda_0) \, \frac{\partial }{\partial  \lambda_0} 
          \right ] m_H^2   \; = \; 0   \; .
\end{equation} 
This last equation shows that $m_H^2$ is  a renormalisation-group invariant physical quantity. Thus, the renormalisation procedure
has introduced an arbitrary dimensional  parameter (dimensional transmutation).  \\
Let us, now, solve Eq.~(\ref{2.21}). We get:
\begin{equation} 
\label{2.22} 
 \beta_{triv}(\lambda_0)  \;= \; - \, a_1  \; \frac{9}{8 \pi^2} \,  \lambda_0^2  \; .
\end{equation}
After that, inserting Eq.~(\ref{2.22}) into Eq.~(\ref{2.20}) we finally obtain:
\begin{equation} 
\label{2.23} 
 \gamma_{triv}(\lambda_0)  \;= \; - \, a_1 \; \frac{9}{16 \pi^2} \,  \lambda_0  \; .
\end{equation}
From Eqs.~(\ref{2.22}) and (\ref{2.23}) we see that:
\begin{equation} 
\label{2.24} 
 \frac{\gamma_{triv}(\lambda)}{  \beta_{triv}(\lambda)}  \; = \; -  \; \frac{1}{2 \, \lambda}   \; .
\end{equation}
A few comments are in order. Firstly,  Eq.~(\ref{2.22}) shows that the beta-function is negative.
As a consequence,  the theory is free asymptotically  in agreement with triviality:
\begin{equation} 
\label{2.25} 
\lim_{ \Lambda \to \infty} \lambda  \;  \; { \large{\sim} } \; \;  \frac{16 \pi^2}{9 a_1} \; \frac{1}{\ln(\frac{\Lambda^2}{m_H^2})}   \;  \; . 
\end{equation}
It is worthwhile to note that the asymptotic behaviour of the self-coupling constant  $\lambda$ does not depend on the bare parameter
$\lambda_0$. 
The triviality of the theory means that the fluctuating quantum field   $\hat{\eta}(x)$ is a free field. On the other hand, a non-zero
value of the anomalous dimension  $\gamma_{triv}$ indicates that the scalar condensate suffers a non-trivial renormalisation. 
Let us introduce the renormalised field $\hat{\eta}_R(x)$ and condensate $\phi_R$. 
Since the fluctuation $\hat{\eta}$ is a free field  with mass $m_H$ we have $\hat{\eta}_R = \hat{\eta}$, namely:
\begin{equation} 
\label{2.26} 
Z_{\eta} \;  =  \; 1 \; . 
\end{equation}
On the other hand, for the scalar condensate  we have:
\begin{equation} 
\label{2.27} 
\phi_R \; = \; Z_{\phi}^{-\frac{1}{2}} \;  \phi  \; \; \; , 
\end{equation}
where, according to Eq.~(\ref{2.23}), we have:
\begin{equation} 
\label{2.28} 
\lim_{ \Lambda \to \infty} \; Z_{\phi}  \;  { \large{\sim} }  \;  \frac{9 a_1}{16 \pi^2 } \;   \ln (\frac{\Lambda^2}{m_H^2})   \;  \; . 
\end{equation}
Finally,  Eq.~(\ref{2.24})  assures that both  $m_H^2 = 3  \tilde{\lambda} \,  v^2$  and  $ \tilde{\omega}^2(\phi) =  3  \tilde{\lambda} \,  \phi^2$
are renormalisation group invariants. This, in turns, shows that  $V_{triv}(\phi)$ , as given by  Eq.~(\ref{2.18}), is manifestly renormalisation group invariant.
To see this,  we note that:
\begin{equation} 
\label{2.29} 
\ln \sqrt  \frac{Z_{\phi}(\mu)}{ Z_{\phi}(\mu_0)}  \; =  \; \int_{\lambda(\mu_0)}^{\lambda(\mu)} 
 \frac{\gamma_{triv}(\lambda')}{\beta_{triv}(\lambda') } \;  d \lambda' \; = \; - \, \frac{1}{2} \, 
\int_{\lambda(\mu_0)}^{\lambda(\mu)} 
  \frac{ d \lambda'}{\lambda'}  \; =  \; - \, \frac{1}{2} \,  \ln \frac{\lambda(\mu)}{\lambda(\mu_0)} \; . 
\end{equation}
Therefore we have:
\begin{equation} 
\label{2.30} 
Z_{\phi}(\mu) \; \lambda(\mu)  \; =  \; Z_{\phi}(\mu_0) \; \lambda(\mu_0)  \; \; . 
\end{equation}
Another  remarkable  consequence  of Eq.~(\ref{2.30}) is that,  after using Eqs.~(\ref{2.25}) and (\ref{2.28}), 
the physical mass $m_H$ is {\it finitely} related to the renormalised vacuum  expectation value of the scalar field  $v_R$:
\begin{equation} 
\label{2.31} 
m_H \;  =  \; \xi \; v_R  \; . 
\end{equation}
It should be clear that the physical mass $m_H$ is an arbitrary parameter of the theory. On the other hand the parameter $\xi$ being a pure number can be determined in the non perturbative lattice approach. \\
The above discussion indicates that the triviality and spontaneous symmetry breaking approach deviates considerably from the perturbative
scheme. In particular, in the triviality proposal the scalar condensate is subject to a non-trivial  rescaling $Z_{\phi}$ which is different
from the wavefunction renormalisation constant $Z_{\eta}$ of the fluctuating field. We suspect that it is this aspect that led to a widespread
skepticism toward the triviality and spontaneous symmetry breaking proposal. The crucial point is that the perturbative approach completely
neglect the quantum-mechanical nature of the vacuum condensation of the scalar field. In fact, we show below that the scalar
condensation is the relativistic version of the Bose-Einstein condensation that, as it is well known, is a purely quantum phenomenon.
As we said, in the perturbative implementation of the vacuum condensation of the quantum scalar field one introduces a tachyonic
mass term that drives the condensation in the vacuum. This process is stabilised by the presence of a repulsive quartic self-coupling.
As a result the vacuum scalar condensate strength $v$ is evaluated by minimising the classical potential. After that, the higher order
corrections are evaluated by means of the perturbative expansion in the small self-coupling. We see, therefore, that it is implicitly 
believed that the vacuum scalar condensation is essentially a semi-classical process. This is not true. Indeed,
let us consider the vacuum expectation value of the quantum scalar field $\hat{\phi}(x)$. Since the Hamiltonian operator annihilates the vacuum,
without loss in generality, we may assume $t=0$. Then:
\begin{equation} 
\label{2.32} 
v  \; = \; < \hat{\phi}(\vec{x}) > \; =\; \frac{1}{V} \;  \left < \int d \vec{x} \;  \hat{\phi}(\vec{x})  \right > \; \; ,
\end{equation}
where $V$ is the spatial volume and we have used the translational invariance of the vacuum. In terms of the Fourier transform 
of the quantum field:

\begin{equation} 
\label{2.33} 
\hat{\phi}(\vec{k})  \; = \;   \int \frac{d \vec{x}}{(2 \pi)^{3/2}}  \; \exp (i \vec{k} \cdot \vec{x}) \;  \hat{\phi}(\vec{x})  \; \; ,
\end{equation}
Eq.~(\ref{2.32}) reads:
\begin{equation} 
\label{2.34} 
 < \hat{\phi}(\vec{k}=0) > \; =   \; \frac{V}{(2 \pi)^{3/2}} \;  v  \;  \; .   
\end{equation}
Therefore, we see that the zero modes of the quantum field  $\hat{\phi}(\vec{k})$  are macroscopically occupied.
In other word the quantum field  $\hat{\phi}$  suffered the Bose-Einstein condensation. Moreover, since the zero-modes $\hat{\phi}(\vec{k}=0)$
become macroscopically occupied,  the creation and annihilation operator $\hat{a}^{\dagger}_{\vec{k}=0}$ and   $\hat{a}_{\vec{k}=0}$
are no longer operators but they can be considered c-numbers. This corresponds exactly to the approximations adopted by N. N. Bogolubov
in his attempt to explain the phenomenon of superfluidity in He II from first principles~\cite{Bogolubov:1947}. We, therefore, may conclude  that
the physical meaning of  Eq.~(\ref{2.2}) is the Bose-Einstein condensation of the relativistic local quantum field  $\hat{\phi}$ in the
Bogolubov's approximations.  It follows that  the Bose-Einstein condensation leads to the following constraints on the quantum field
$\hat{\phi}(\vec{k})$:
\begin{equation} 
\label{2.35} 
\left . \hat{\phi}(\vec{k}) \right |_{\vec{k}=0}  \; =  \;  \frac{V}{(2 \pi)^{3/2}} \;  v  \;  \; .   
\end{equation}
We are led to a scalar quantum field with Dirichlet boundary conditions on the hypersurface $\vec{k}=0$. In a classical paper by K. Symanzik
on the Schr\"odinger representation in quantum field theories~\cite{Symanzik:1981} it was demonstrated that the field operator that is being
diagonalised on a smooth hypersurface differs from the usual renormalised field by a diverging factor. In our case the divergency related to
the presence of the boundary condition can be eliminated by adding one more counterterm to the usual counterterms. More precisely,
we need to renormalise the zero-mode quantum field $ \hat{\phi}(\vec{k}) |_{\vec{k}=0} $ with a renormalisation constant ($Z_{\phi}$) that
is different from the usual wavefunction renormalisation constant ($Z_{\eta}$) of the fluctuating fields $\hat{\phi}(\vec{k}) , \vec{k} \neq 0$.
Another important aspect to be considered is that the creation and annihilation operators of the vacuum with condensate are
related to the ones of the vacuum without condensate by the Bogolubov's canonical transformations ~\cite{Bogolubov:1947}. As a
consequence, the two vacua are inequivalent in accordance with the Haag's theorem. The usually adopted perturbative scheme,
by considering the vacuum scalar condensation as a quasi-classical phenomenon, completely overlooks these important aspects.
Therefore, we see that the renormalised perturbative approach could lead to misleading results.\\
Let us summarise the main results of this rather technical Section. Basically, what we have shown is that a free massless scalar quantum
field in presence of an even infinitesimal positive quartic self-coupling is unstable towards the phase with Bose-Einstein condensation.
As a result, the stable phase is a massive free scalar field and a finite scalar vacuum condensate where the scalar field mass is
finitely related to the vacuum condensate strength. \\
The problem is that, within the pure scalar sector, we do not have a physical mechanism to generate a positive self-coupling needed
to trigger the spontaneous symmetry breaking mechanism. So that, the scalar sector of the Standard Model being trivial cannot give
rise to the Bose-Einstein condensation. If this is the case, we should have free massless scalar particles.  However, in Sect.~\ref{s-4}
we will argue that, once the Higgs scalar fields are coupled to the gauge vector bosons of the Standard Model, then there are
well-defined physical processes able to trigger the Higgs mechanism.
\section{Comparison with non-perturbative numerical simulations}
\label{s-3}
In the previous Section we have discussed in details two different approaches on  spontaneous symmetry breaking for the one-component
quantum scalar field theory. We have seen that these two approaches lead to sizeable differences that, in principle, can be tested 
by comparison with a truly non-perturbative approach to quantum scalar theories. Actually, the lattice approach to quantum field theories 
 offers us the unique opportunity to study a quantum field theory  by means of non-perturbative methods. 
 To  this end, firstly one needs to formulate
 the given quantum field theory in the Euclidean space by performing the analytic continuation to imaginary time. The technical convenience of considering field theories at purely imaginary times resides on the possibility to study a given quantum field theory by means of non-perturbative
numerical  techniques. In this respect it should be noticed that Osterwalder and Schrader established the complete equivalence between relativistic
 and Euclidean quantum field theories~\cite{Osterwalder:1973,Osterwalder:1975}. In particular, the Osterwalder-Schrader positivity
 (also called reflection positivity) allows the reconstruction of the relativistic quantum theory from the Euclidean correlation functions (Schwinger
 function) and leads to a correspondence between the physical states of the real-time relativistic theory and the states obtained by acting with the
 Euclidean operators on the quantum vacuum. \\
 Starting from the classical Lagrangian Eq.(\ref{2.1}) one readily obtains the Euclidean action:
\begin{equation} 
\label{3.1} 
S_E = \int d^4 x \;  {\cal{L}}_E(x) \;  = \; \int d^4 x  \left [ \frac{1}{2} \, (\partial_\mu \hat{\phi})^2 \; + \; \frac{1}{2} \, \mu^2  \,  \hat{\phi}^2 \; + \;
 \frac{1}{4} \,  \lambda_0  \,  \hat{\phi}^4  \right ] \; .
\end{equation}
 The Euclidean quantum field dynamics is dealt with  from  the  functional integral point of view.  In  the lattice approach the Euclidean
four-dimensional space-time  is discretised to a hypercubic lattice with finite lattice spacing $a$,  so that, in principle, the Schwinger functions can be
evaluated by means of numerical functional integrations. \\
The lattice version of the  Euclidean action  Eq.~(\ref{3.1}) is written as:
\begin{equation}
\label{3.2}
  S _E \; = \; a^4 \,  \sum_x \left [ \frac{1}{2 a^2}  \sum_{\mu=1}^{4}  \left ( \phi(x+a \hat{\mu}) - 
\phi(x) \right )^2 \; + \;  
\frac{r_0}{2} \, \phi^2(x) \;  + \; 
\frac{\lambda_0}{4} \, \phi^4(x)  \right ]   \; , 
\end{equation}
where $x$ stands for a generic lattice site,  and $r_0$ and $\lambda_0 >0$ are free parameters.  It is customary to adopt
lattice units where $a = 1$. \\
In general, the continuum limit of the lattice theory $a \rightarrow 0$ is obtained at a critical point
of the lattice system  where the dimensionless correlation length diverges.  It turned out 
that~\footnote{A good account can be found in Ref.~\cite{Lang:1994}}, for each given value of $\lambda_0$, there
exists a critical value $r_c(\lambda_0)$ such that the system is in the symmetric phase (vacuum expectation value of $\phi$ vanishes)
for $r_0 > r_c(\lambda_0)$, while it is in the broken phase (non-zero vacuum expectation value of $\phi$) for $r_0 < r_c(\lambda_0)$.
Remarkably, irrespective to the value of $\lambda_0$, the numerical simulations of the four-dimensional one-component scalar field theory on the 
lattice indicated that the continuum limit defines a trivial theory  in both the symmetric and broken phases, in accordance
with the recent rigorous result~\cite{Aizenman:2021}.  In other words, the continuous limit of the lattice theory is governed by the
Gaussian fixed point $\lambda^* = 0$. Therefore, at the critical point of the model the critical exponents that characterise the 
singular behaviour of various quantities take the four-dimensional mean field values.  In general,
at a critical point the correlation length is much higher than the lattice spacing,  so that one may send the lattice spacing to zero 
while keeping a fixed value for the physical mass parameter. In the scaling region, once the scale is set by the lattice spacing, 
all other parameters may then be expressed in terms of this scale parameter.  Note, however, that  all other
physical parameters should scale correspondingly such that one can  identify the correct continuum theory.
This scaling regime  may be observed quite early in the parameter space. Practically,  the scaling region
is characterised by a correlation length slightly higher than the lattice spacing. On the other hand,  one has to be quite close to the critical
point in order to observe the asymptotic scaling, namely how the lattice spacing tends to zero when $ \lambda \rightarrow  \lambda^*$.
This asymptotic scaling is determined  from the renormalisation group $\beta$-function. \\
The generally accepted assumption is that the renormalised perturbation theory  is required to define the continuum limit. In this case,
the asymptotic scaling is set by  the two-loop perturbative $\beta$-function:
\begin{equation} 
\label{3.3} 
 \beta_{pert}(\lambda) \;= \; \frac{9}{8 \pi^2} \,  \lambda^2 \; - \;   \frac{51}{64 \pi^4} \,  \lambda^3 \; .
\end{equation}
Accordingly, the asymptotic scaling is governed by:
\begin{equation} 
\label{3.4} 
 ( a \, m)_{pert} \; \sim   ( \frac{9}{8 \pi^2} \,  \lambda )^{\frac{17}{27}}  \;  \exp [ - \,  \frac{8 \pi^2}{9}\,  \frac{1}{\lambda}] \; .
\end{equation}
This scaling law is assumed to hold both in the symmetric phase~\cite{Luscher:1987} and in the broken 
phase~\cite{Luscher:1988}~\footnote{Note that the coupling constant $g$ in the L\"uscher and Weisz scheme is related
to our $\lambda$ by $g = 6 \,  \lambda$.}. From Eq.~(\ref{3.4}) we see that  in the continuum limit $am \rightarrow 0$  the coupling constant
$\lambda$ is forced to go to zero. However,   the ultraviolet cutoff $\sim 1/a$  cannot be pushed to arbitrary high values due to the presence
of a singularity  (the Landau pole). Thus, we must admit that the continuum limit corresponds to $m \rightarrow 0$, i.e. the perturbative
approach describes the triviality of the theory as an infrared Gaussian fixed point. Therefore, taking these results at face value, for fixed $\lambda$ 
only at finite cutoff  a non-vanishing value of $\lambda$  is permitted. In this way within perturbation theory even a trivial theory may have 
an effective interaction below the cutoff.  \\
On the other hand, in the  broken phase the triviality and spontaneous symmetry breaking picture leads, according to
Eq.~(\ref{2.22}), to the asymptotic scaling law:
\begin{equation} 
\label{3.5} 
 ( a \, m)_{triv} \; \sim   \;  \exp[ - \,  \frac{8 \pi^2}{9 a_1} \, \frac{1}{ \lambda}] \; .
\end{equation}
This last equation shows that the continuous limit is an ultraviolet Gaussian fixed point leading to a free massive scalar field with mass
finitely related to the condensate strength. \\
In principle, the two different approaches to the spontaneous symmetry breaking for the one-component scalar field can be
checked non perturbatively by  direct lattice numerical simulations in a region of the coupling parameter space where
 the asymptotic scaling associated with the trivial  Gaussian fixed point sets in.  The first attempt in this direction was advanced in
 Refs.~\cite{Agodi:1997,Agodi:1998} where it was numerically computed the effective potential on the lattice by employing
 the Euclidean action Eq.~(\ref{3.2}) with $\lambda_0 = 0.5$. The relevant lattice observables were  the  vacuum expectation value 
 of the scalar field (also called the magnetisation):
\begin{equation}
\label{3.6}
 v \; = \; \langle |\phi| \rangle \quad , \quad \phi  \equiv  \frac{1}{L^4} \; \sum_x \phi(x) \; ,
\end{equation}
and the  zero-momentum susceptibility:
\begin{equation}
\label{3.7}
 \chi \; = \; L^4 \left[ \left\langle |\phi|^2
\right\rangle - \left\langle |\phi| \right\rangle^2 \right]  
\end{equation}
where $L$ is the lattice linear size.
%
%
\begin{figure}
\vspace{-1.0cm}
\begin{center}
\includegraphics[width=0.7\textwidth,clip]{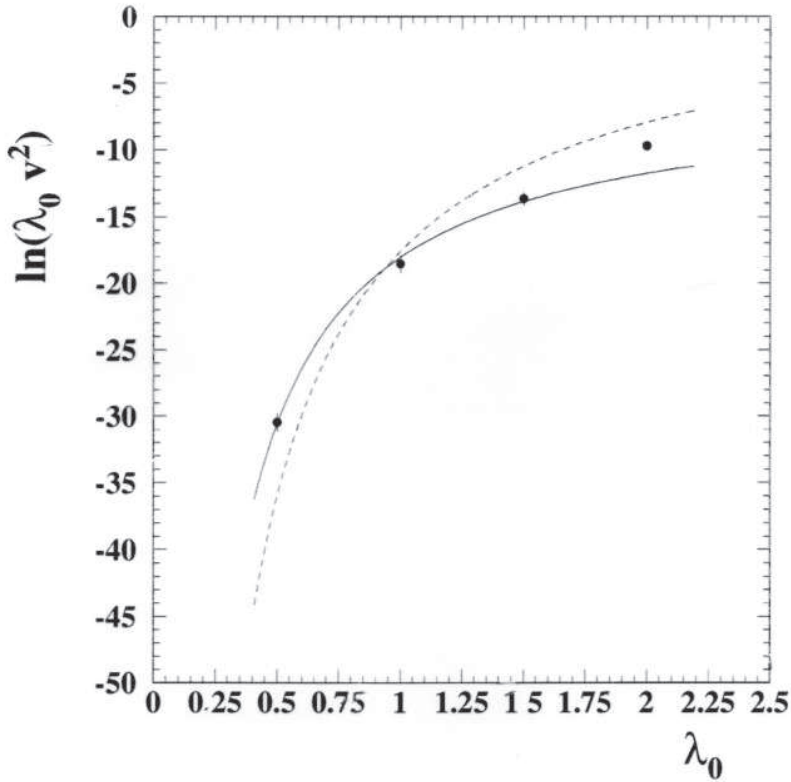}
\end{center}
\vspace{-0.5cm}
\caption{\label{Fig1} 
The lattice data for $ \ln ( \lambda_0 v^2 )$ for different values of $\lambda_0$.
The continuous line corresponds to the best fit with Eq.~(\ref{3.9}), while the dashed line corresponds to the two-loop fit Eq.~(\ref{3.8}).
}
\end{figure}
The effective potential  can be analysed  in the lattice  formulation by computing the vacuum expectation value of the field as a function of an
 uniform external source with strength J.   Determining $v(J)$ at several J-values is equivalent  to  invert the relation
$ J = d V_{eff}/d v$ involving the effective potential $V_{eff}(v)$.  
As matter of fact, the one-loop potential agrees remarkably well with the lattice results, while its perturbative improvement  fails to
 reproduce the Monte Carlo data. More precisely, in  Refs.~\cite{Agodi:1997,Agodi:1998} it was compared the two-loop
renormalisation-group improved effective potential with the one in the triviality and spontaneous symmetry breaking picture
discussed at length in Sec.~\ref{s-2}. The numerical data significantly favour the unconventional interpretation
of triviality  over the conventional perturbative interpretation. However, it should  be desirable to have a more direct test
of the two asymptotic scaling laws. To this end, we note that in the critical region $ m^2 \sim \lambda_0 v^2$. Therefore,
according to Eqs.~(\ref{3.4}) and (\ref{3.5}) we can write:
\begin{equation} 
\label{3.8} 
 \left [ \log ( \lambda_0 \, v^2) \right ]_{pert} \; \sim \; - \,  \frac{16 \pi^2}{9} \,  \frac{1}{\lambda}   \; + \;
\frac{34}{27}\,  \log ( \frac{9}{8 \pi^2} \,  \lambda ) \; + C_1 \; ,
\end{equation}
\begin{equation} 
\label{3.9} 
 \left [ \log ( \lambda_0 \, v^2) \right ]_{triv} \; \sim \;  -   \;   \frac{16 \pi^2}{9 a_1} \, \frac{1}{ \lambda}   \; + C_2 \; 
\end{equation}
$C_1$, $C_2$ being some constant.  Indeed,  having at our disposal the vacuum expectation value of the scalar field for four different values 
of the coupling constant in the range $0.5 \leq \lambda_0 \leq 2.0$~\footnote{These results are based on a talk given by the author at the 15th International Symposium on Lattice Field Theory (Lattice 1997), Edinburgh, UK, July 22 - 26, 1997.}, we tried to compare the two asymptotic
scaling laws.  In Fig.~\ref{Fig1} we display the best fits to the numerical data with both Eqs.~(\ref{3.8}) and (\ref{3.9}). Looking at 
Fig.~\ref{Fig1} it is evident  that the triviality scaling law compare to the data better than the two-loop scaling law. From the fit to the 
numerical data we inferred that:
\begin{equation} 
\label{3.10} 
 a_1 \; = \;  1.18(7) \; \; . 
\end{equation}
Even thought the above results are promising, it should be mentioned that at the  the upper critical dimension (D= 4)  the
 leading scaling behaviour for self-interacting scalar theory is coincident with mean field theory but, however, there are 
 multiplicative logarithmic corrections that depend on the asymptotic scaling law. Unfortunately, for $\lambda_0 \sim 1$
the numerical data do not show evidence of these logarithmic corrections (see Refs.~\cite{Agodi:1997,Agodi:1998}).
As a matter of fact, for a model-independent check of the asymptotic scaling law it would be desirable to have
numerical data that display as clear as possible the effects of the logarithmic corrections to the mean-field scaling
behaviour. Evidently, the logarithmic corrections are related to the non-Gaussian quantum fluctuations that, in turns,
depend on the strength of the quartic self-coupling $\lambda_0$. For this reason it is   customary to perform numerical simulations
 in the so-called Ising limit. \\
The Ising limit corresponds to $\lambda_0 \; \rightarrow \; \infty$. It easy to see that,
in this limit, the one-component scalar field theory becomes governed by the lattice action (see, eg, Ref.~\cite{Luscher:1987}):
\begin{equation}
\label{3.11}
S_{\text{Ising}} \;  = \; - \, \kappa  \, \sum_x\sum_{\mu} \,  \left [  \phi(x+\hat{\mu}) \phi(x) \, + \, \phi(x-\hat{\mu}) \phi(x)  \right ]
\end{equation}
where $\phi(x)$ takes only the  values $+1$ or $-1$.  
It is known that there is a critical coupling~\cite{Gaunt:1979}:
\begin{equation}
\label{3.12}
\kappa_c \; = \;   0.074834(15) \; 
 \end{equation}
such that for $\kappa > \kappa_c$ the theory is in the  broken phase, while for  $\kappa <  \kappa_c$ it
is in the symmetric phase. The continuum limit corresponds to $\kappa \rightarrow \kappa_c$ where $a m \rightarrow 0$. \\
As discussed in  Section~\ref{s-2}, the triviality of the scalar theory means that the renormalised self coupling vanishes as   $\frac{1}{\ln(\frac{\Lambda^2}{m_H^2})}$ when   $\Lambda \rightarrow \infty$. Since  on the lattice the ultraviolet cutoff is 
$\Lambda = \frac{\pi}{a}$,  we have:
\begin{equation} 
\label{3.13} 
 \lambda \; \sim  \; \frac{1}{ \ln(\frac{\Lambda}{m_H})}  \; \sim  \; \frac{1}{ \ln(\frac{\pi}{a m})} \;  \; . 
\end{equation}
We have seen that the perturbative interpretation of triviality assumes that in the continuum limit there is an infrared  Gaussian fixed point.
On the other hand,  according to Section~\ref{s-2},  in the triviality and spontaneous symmetry breaking scenario the continuum dynamics is governed by an  ultraviolet Gaussian fixed point.  As we discuss below, these two different interpretation of triviality lead to different logarithmic correction to the Gaussian scaling laws that can be checked with numerical simulations on the lattice. \\
\indent
In  Ref.~\cite{Cea:2005,Cea:2012} extensive numerical lattice simulations of the  one-component scalar field theory in the Ising limit 
have been  performed using the Swendsen-Wang~\cite{Swendsen:1987}  and Wolff~\cite{Wolff:1989} cluster algorithms.
According  to the perturbative scheme  one expects:
\begin{equation}
\label{3.14}
 \lim_{ \kappa \rightarrow \kappa_c^{+}}    v^2 \;  \chi  \; \sim \; 
| \ln(\kappa-\kappa_c)| \; .
\end{equation}
On the other hand, since in the triviality and spontaneous symmetry breaking scenario one expects that
$Z_\varphi  \sim \ln(\frac{\Lambda}{m_H})  \sim|\ln(\kappa-\kappa_c)| $ we should have:
\begin{equation}
\label{3.15}
 \lim_{ \kappa \rightarrow \kappa_c^{+}}     v^2 \;  \chi \;  \sim \;  |\ln(\kappa-\kappa_c)|^2   \; .
\end{equation}
The prediction in Eq.~(\ref{3.15})  can be directly compared with the lattice data  reported in  Ref.~\cite{Cea:2005,Cea:2012}.
Performing a fit to the lattice data with the functional form:
\begin{equation}
\label{3.16}
 v^2 \;  \chi  \; = \; K \,  |\ln(\kappa-\kappa_c)|^2 \; ,
\end{equation}
one  obtains a rather good fit of the lattice data with~\cite{Cea:2012}:
\begin{equation}
\label{3.17}
 K \; = \;  0.07560(49)  \; \; , \; \;   \kappa_c \; = \; 0.074821(12) \; \; , \; \; \chi^2/dof  \sim   1.0 \; .
\end{equation}
\begin{figure}
\begin{center}
\vspace{-1.0cm}
\includegraphics[width=0.7\textwidth,clip]{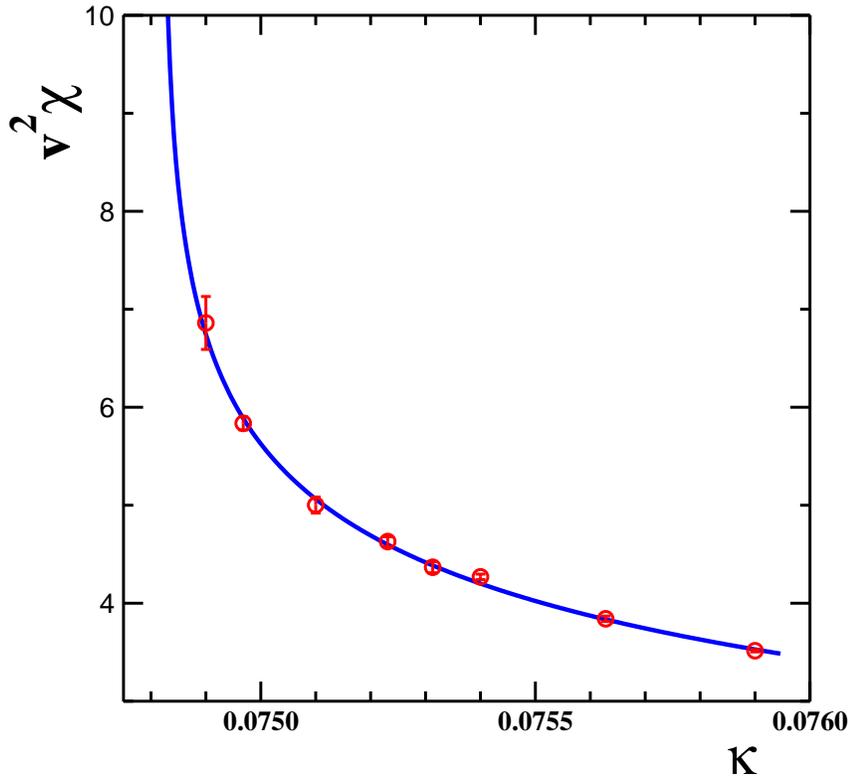}
\end{center}
\vspace{-0.5cm}
\caption{\label{Fig2} 
(Color online) Lattice data for $v^2 \chi$ versus $\kappa$  together with the  fit Eq.~(\ref{3.16}) (solid line).
}
\end{figure}
The results of the fit are  displayed in Fig.~\ref{Fig2}.  Note that  the precise determination of the critical coupling $\kappa_c$ in Eq.~(\ref{3.17})
 is in good agreement with  Eq.~(\ref{3.12}). \\
On the other hand,  the prediction based on 2-loop renormalised perturbation theory can be written as~\cite{Balog:2004}:

\begin{equation}
\label{3.18}
      \left[  v^2 \;  \chi  \right]_{pert} \; = \;  A \, ( {l} - \frac{25}{27} \ln { l}) \; + \;  B 
\end{equation}
where $l=|\ln (\kappa -\kappa_c)|$) and the constants $A$ and $B$ are  constrained by  the theoretical relations~\cite{Luscher:1987,Luscher:1988}:
\begin{equation}
\label{3.19}
A \; = \; 1.20(3) \; \; \; \; , \; \; \;  B  \; = \; - \, 1.6(5) \; \; .
\end{equation}
If one try to fit the data with Eq.~(\ref{3.18}) taking into account the constrains Eq.~(\ref{3.19}), then  the resulting fit is rather poor
(see the dashed line in Figure~1 of Ref.~\cite{Cea:2012}):
\begin{equation}
\label{3.20}
  A \; = \;  1.17  \; , \;   B  \; = \; - \,  2.10  \; \; , \; \; \kappa_c \; = \; 0.074800(1) \; \; , \; \; \chi^2/dof \sim  10^2 \; .
\end{equation}
It should be mentioned that the above results have been subject to an intense debate in the literature
(see, for instance, Refs.~\cite{Stevenson:2005,Balog:2006}  and references therein).
\\
Aside from the previous evidences in favour of the triviality and spontaneous symmetry breaking proposal, further  evidences would come 
from the direct detection of the increases of the condensate rescaling $Z_\phi   \sim |\ln(\kappa-\kappa_c)| $. Indeed, observing that:
\begin{equation}
\label{3.21}
  Z_\phi  \;  =   \;   2 \; \kappa \,  m^2  \; \chi  \; ,
\end{equation}
one can easily extract the condensate rescaling  $Z_\phi$ from the available lattice measurements.
Here, in Fig.~\ref{Fig3} we report  the lattice data  obtained in Ref.~\cite{Cea:2005,Cea:2012}  for $Z_\phi$, as defined in
Eq.~(\ref{3.21}),  versus $a m$  reported in  Ref.~\cite{Luscher:1988}. 
 For comparison we also report the wavefunction renormalisation constant of the fluctuating field $Z_{\eta}$ taken from  
Ref.~\cite{Luscher:1988}.  The lattice data are perfectly consistent with the logarithmic increases  
 $Z_\phi \;   \sim  \;   \ln ( \, \frac{1}{ a m})$ (see the dashed line  in Fig.~\ref{Fig3}).
\begin{figure}
\begin{center}
\vspace{-1.49cm}
\includegraphics[width=0.7\textwidth,clip]{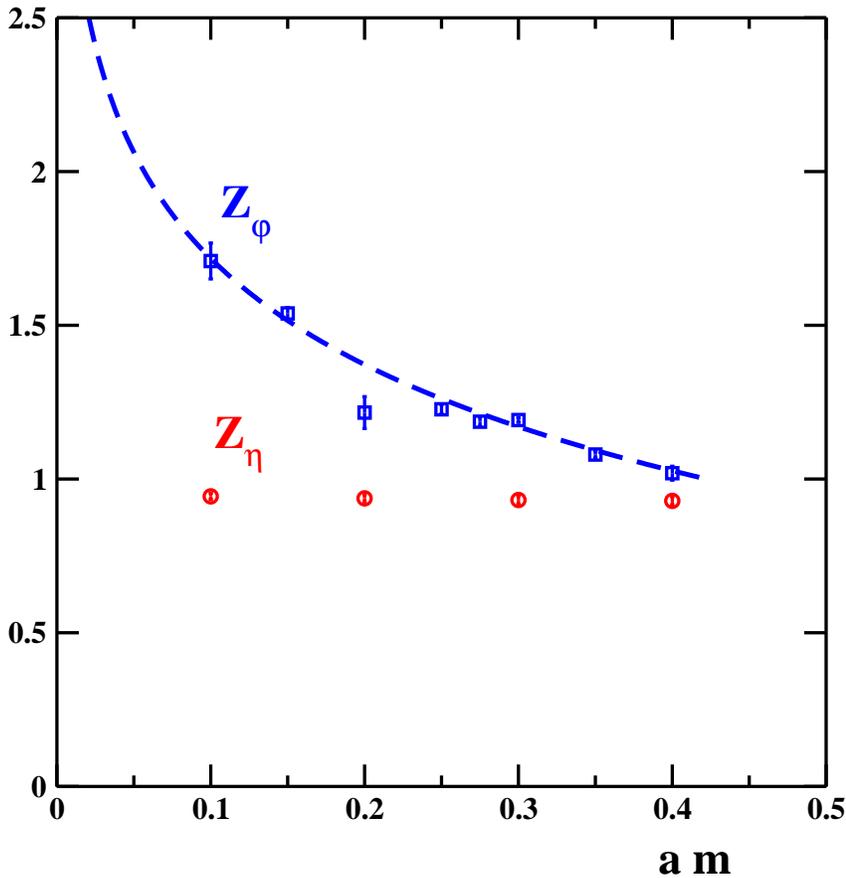}
\end{center}
\vspace{-0.7cm}
\caption{\label{Fig3} 
(Color online) The values of  condensate rescaling $Z_\phi$ and the wavefunction renormalisation constant  $Z_\eta$ versus $a m$.
The figure has been adapted from Fig.~2 of Ref.~\cite{Cea:2012}.
}
\end{figure}
These evidences for the non-trivial rescaling of the vacuum condensate has been criticised by the authors of Ref.~\cite{Balog:2004}.
In particular these authors claim that  there are no evidences  of a rescaling of the vacuum condensate that is different from
the  wavefunction  renormalisation  constant. However, they performed numerical simulations at two finite values of the
quartic self-coupling $\lambda_0 \sim 1$. We already noticed that for   $\lambda_0 \sim 1$ there are no numerical evidences of the needed
logarithmic corrections to the Gaussian mean field scaling laws. On the other hand, in the unique simulation at the Ising limit 
they measured the propagator in momentum space with   lattice momenta  $\hat{p}^2 $, where  $\hat{p}_\mu =2 \sin p_\mu/2$.
Previous studies indicated that  the effects of the logarithmic increase of   $Z_\phi$ on the connected propagator manifest itself only at 
very small momenta  $\hat{p}^2   \lesssim  0.1$ (compare  Fig.~3 with  Fig.~4  in  Ref.~\cite{Cea:2005}).  We suspect that the authors of 
Ref.~\cite{Balog:2004} do not reach  values of momenta  very close to zero. Indeed, the wavefunction renormalisation constant
reported in Table 5 of Ref.~\cite{Balog:2004} agrees with the wavefunction renormalisation constant $Z_\eta$ displayed in Fig.~\ref{Fig3}. \\
In the previous Section we have seen that, by adopting the alternative interpretation of triviality,  the mass of the  excitations over the vacuum
condensate behave like a massive scalar field with mass $m_H$  related to the renormalised vacuum condensate by Eq.~(\ref{2.31}).
Using the lattice data one may easily estimate the numerical constant $\xi$ in Eq.~(\ref{2.31}). Indeed, in Fig.~\ref{Fig4}, following
the analyses reported in Ref.~\cite{Cea:2012},  we display  $m_H$ versus $a \, m$ together  with our estimate of the extrapolated continuum
limit obtained by assuming for the renormalised vacuum condensate the Standard Model value $v \simeq 246$ GeV: 
\begin{equation}
\label{3.22}
m_H \; =750 \; \pm 30 \;   \; {\text{GeV}} \; ,
\end{equation}
where the displayed  error includes both the statistical and systematic  errors.  
\begin{figure}
\begin{center}
\vspace{-0.5cm}
\includegraphics[width=0.8\textwidth,clip]{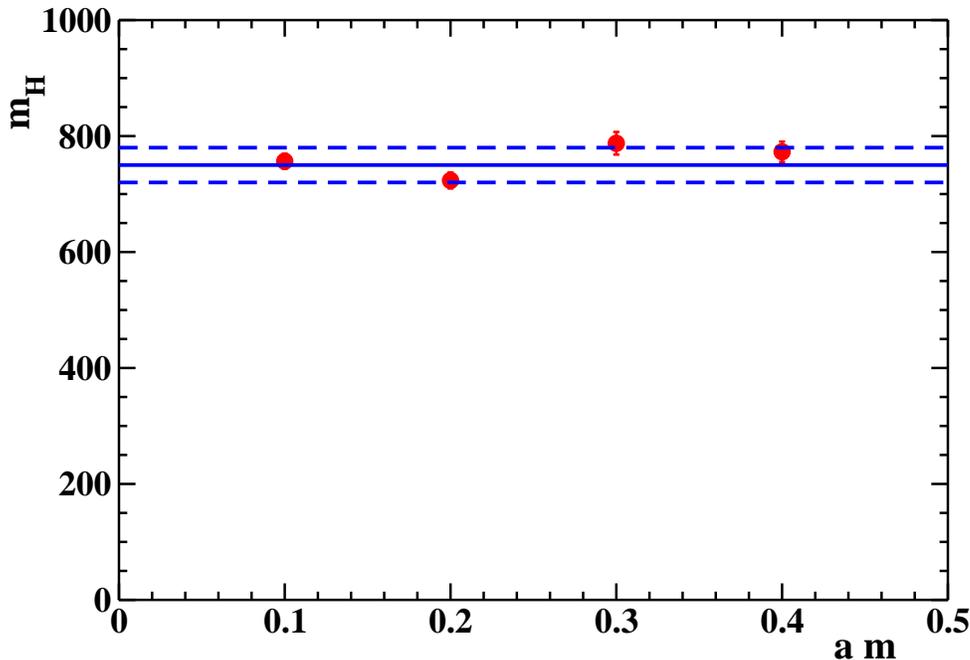}
\end{center}
\vspace{-0.5cm}
\caption{\label{Fig4} 
(Color online) The values of $m_H$ obtained from the lattice data in Ref.~\cite{Cea:2012} versus
$a \, m$ assuming $v \; \simeq \; 246$ GeV. The horizontal lines  correspond  to  Eq.~(\ref{3.22}). 
}
\end{figure}
In the next Section we shall see that  the  lattice estimate of the Higgs mass Eq.~(\ref{3.22}) is  relevant for the physical Higgs boson
of the Standard Model. 
\\
A further check of the asymptotic scaling laws can be obtained by means of
the finite-size scaling theory derived using the renormalisation group
methods~\cite{Brezin:1976,Zinn-Justine:2002}. In the following we shall follow
closely Ref.~\cite{Kenna:1993} where it is derived a finite-size scaling theory for the 
 one-component scalar field theory in four dimensions. \\
Let us consider the partition functional  $Z[H,t]$ in presence of external sources:
\begin{equation}
\label{3.23}
S_{\text{Source}} \;  = \;  - \;  \sum_x   \,  \left [  H(x) \; \phi(x)  \;  + \;  \frac{1}{2} \; t(x) \, \phi^2(x)   \right ] \;.
\end{equation}
Writing:
\begin{equation}
\label{3.24}
Z[H,t]   \;  = \;   \exp \{ \,  W[H,t] \, \}  \;  , \;  W[0,0] \; = \; 0 \; , 
\end{equation}
by a Legendre transformation of $W[H,t]$ with respect to the external field $H(x)$  one introduces the generating functional
 $\Gamma[M,t]$ where:
\begin{equation}
\label{3.25}
M(x)   \;  = \;   \frac{\delta \, W[M,t]}{\delta \, H(x)}  \; .
\end{equation}
The connected correlation functions can be obtained by functional derivatives of the generating functional  $\Gamma[M,t]$:
\begin{equation}
\label{3.26}
\Gamma^{(K,N)}[y_1,...,y_K,x_1,...,x_N; M,t] \; = \; 
  \frac{\delta^{K+N} \, \Gamma[M,t]}{\delta t(y_1) ... \delta t(y_K)  \, \delta M(x_1) ... \delta M(x_N)} \; .
\end{equation}
After renormalisation the renormalised connected correlation functions satisfy the renormalisation group equations that express the invariance 
of the physics under a rescaling $\lambda$ of the mass parameter~\cite{Brezin:1976}.
This allows one to examine the approach to criticality from within both the symmetric and the broken phases. \\
We known that in the renormalised perturbation approach the critical point is a Gaussian infrared fixed point. The relevant flow equations are:
\begin{eqnarray} 
\label{3.27} 
\nonumber
\lim_{ \lambda \to 0 } \; \frac{g_R(\lambda)}{g_R}    \; { \large{\sim} } \;   \frac{2}{3 g_R} \; \frac{1}{ | \ln \lambda | }   \;  \;  \; \;  \;  \; \;   pert.
\\
\lim_{ \lambda \to 0 } \; \frac{t(\lambda)}{t}    \; { \large{\sim} } \;  \left [  \frac{2}{3} \; \frac{1}{ | \ln \lambda | }  \right ]^{\frac{1}{3}}  \;  \;  \; \;  \;  \; \;   pert.
 \\
 \nonumber
\lim_{ \lambda \to 0 } \; \frac{M(\lambda)}{M}    \; { \large{\sim} } \;  1   \;  \;  \; \;  \;  \; \;   pert.
\end{eqnarray}
where  $g_R(1)=g_R$, $t(1)=t$,  $M(1)=M$. On the other hand, in the triviality and spontaneous symmetry breaking picture there is a ultraviolet 
 Gaussian fixed point. In this case the flow equations are~\footnote{In perturbation theory the quantum phase transition at the critical point is
 continuous. In the triviality and spontaneous symmetry breaking picture the order of the transition is not known. The phase transition could well be
  first order. However,  numerical simulations on the lattice indicated that the phase transition is almost continuos. In other words, if the transition is first order, then  it must be  a  weak first-order phase transition.}:
\begin{eqnarray} 
\label{3.28} 
\nonumber
\lim_{ \lambda \to 0 } \;\frac{ g_R(\lambda)}{g_R}    \; { \large{\sim} } \;   \frac{2}{3 \, a_1 g_R} \; \frac{1}{ | \ln \lambda | }   \;  \;  \; \;  \;  \; \;   triv.
\\
\lim_{ \lambda \to 0 } \; \frac{t(\lambda)}{t}    \; { \large{\sim} } \;  \left [  \frac{2}{3 \, a_1} 
\; \frac{1}{ | \ln \lambda | }  \right ]^{\frac{1}{3 \, a_1}}  \;  \;  \; \;  \;  \; \;   triv.
 \\
 \nonumber
\lim_{ \lambda \to 0 } \; \frac{M(\lambda)}{M}    \; { \large{\sim} } \;  1   \;  \;  \; \;  \;  \; \;   triv.
\end{eqnarray}
The renormalisation constants of the infinite-volume theory render finite the finite-volume theory too. As a consequence the renormalisation group
equations for the finite-volume theory are the same as in the case of infinite volume. This allows to develop a  finite-size scaling theory for the 
model under study. Actually, we are interested in the correlation function $\Gamma^{(0,0)}_R$. In the renormalised perturbation theory
one readily obtain~\cite{Kenna:1993}: 
\begin{equation}
\label{3.29}
\Gamma^{(0,0)}_R \; \sim \; \Gamma^{(0,0)}_R \left [ L^2 \, t \, (\frac{2}{3 g_R \ln L})^{\frac{1}{3}}, L \, M, \frac{2}{3 \ln L} \right ] 
 \; + \; \frac{3}{4} (\frac{2}{3 g_R})^{\frac{2}{3}} \,  t^2 (\ln L)^{\frac{1}{3}}
 \;  \; \;   \;  \;   \;  \;   \;  \;   \;  \; pert. 
\end{equation}
where $L$ is the linear extent of the system. In the triviality picture we have:
\begin{equation}
\label{3.30}
\Gamma^{(0,0)}_R \; \sim \; \Gamma^{(0,0)}_R \left [ L^2 \, t \, (\frac{2}{3 a_1 g_R \ln L})^{\frac{1}{3 a_1}}, L \, M, \frac{2}{3 a_1 \ln L} \right ] 
 \; + \; \frac{3 a_1}{4} (\frac{2}{3 a_1 g_R})^{\frac{2}{3 a_1}} \,  t^2 (\ln L)^{1 - \frac{2}{3 a_1}}
 \;  \;   \;  \;   triv. 
\end{equation}
Since in both approaches the renormalised coupling is small, perturbation theory may be applied to calculate $\Gamma^{(0,0)}_R$:
\begin{equation}
\label{3.31}
\Gamma^{(0,0)}_R \; \simeq  \;  \left [  g_R \,  M^4 \; + \; t  \, M^2  \right ]  \; ,
\end{equation}
so that:
\begin{equation}
\label{3.32}
\Gamma^{(0,0)}_R \; \sim \; c_1 \;  t \, M^2 \;  ( \ln L)^{ - \frac{1}{3}} \; + \; c_2 \,  M^4 \,  ( \ln L)^{- 1}  + \; c_3 \,  t^2 \,  ( \ln L)^{ \frac{1}{3}}
  \;  \; \;   \;  \;   \;  \;   \;  \;   \;  \; pert. 
\end{equation}
\begin{equation}
\label{3.33}
\Gamma^{(0,0)}_R \; \sim \; c_1 \;  t \, M^2 \;  ( \ln L)^{ - \frac{1}{3 a_1}} \; + \; c_2 \,  M^4 \,  ( \ln L)^{- 1}  + \; c_3 \,  t^2 \,  ( \ln L)^{1 -  \frac{2}{3 a_1}}
  \;  \; \;   \;  \;    triv. 
\end{equation}
Since
\begin{equation}
\label{3.34}
H(x)   \;  = \;   \frac{\delta \, \Gamma[M,t]}{\delta \, M(x)}  \; \;  ,
\end{equation}
we readily obtain:
\begin{equation}
\label{3.35}
H \; \simeq \; c_4 \;  t \, M  \;  ( \ln L)^{ - \frac{1}{3}} \; + \; c_5  \,  M^3 \,  ( \ln L)^{- 1}  
  \;  \; \;   \;  \;   \;  \;   \;  \;   \;  \; pert. 
\end{equation}
\begin{equation}
\label{3.36}
H \; \simeq \; c_4 \;  t \, M  \;  ( \ln L)^{ - \frac{1}{3 a_1}} \; + \; c_5  \,  M^3 \,  ( \ln L)^{- 1}  
  \;  \; \;   \;  \;   \;  \;   \;  \;   \;  \;triv. 
\end{equation}
So that the free energy density in presence of an external field is:
\begin{equation}
\label{3.37}
W_L(t,H) \; \simeq \; c_1^{'}  \;  t \, M^2   \;  ( \ln L)^{ - \frac{1}{3}} \; + \; c_2^{'}  \,  M^4 \,  ( \ln L)^{- 1}  \; + \; c_3  \,  t^2 \,  ( \ln L)^{\frac{1}{3}}  
  \;  \; \;   \;  \;   \;  \;   \;  \;   \;  \;    \;  \;   \;  \;   pert. 
\end{equation}
\begin{equation}
\label{3.38}
W_L(t,H) \; \simeq \; c_1^{'}  \;  t \, M^2   \;  ( \ln L)^{ - \frac{1}{3 a_1}} \; + \; c_2^{'}  \,  M^4 \,  ( \ln L)^{- 1}  \; + \; c_3  \,  t^2 \,  
( \ln L)^{1 - \frac{1}{3 a_1}}  
  \;  \; \;   \;  \;   \;  \;   \;   \;triv. 
\end{equation}
where $M$ is related to $H$ by Eqs.~(\ref{3.35}) and (\ref{3.36}) respectively. \\
If $H=0$, from Eqs.~(\ref{3.35}) and (\ref{3.36}) we infer:
\begin{equation}
\label{3.39}
M^2 \; \simeq \; t \,  ( \ln L)^{ \frac{2}{3}} \;   
  \;  \; \;   \;  \;   \;  \;   \;  \;   \;  \;  \; \; pert. 
\end{equation}
and 
\begin{equation}
\label{3.40}
M^2 \; \simeq \; t \,  ( \ln L)^{1 - \frac{1}{3 a_1}} \;  
  \;  \; \;   \;  \;   \;  \;   \;   \; triv. 
\end{equation}
Thus:
\begin{equation}
\label{3.41}
W_L \; \simeq \; t^2 \,  ( \ln L)^{ \frac{1}{3}} \;   
  \;  \; \;   \;  \;   \;  \;   \;  \;   \;  \;  \; \; pert. 
\end{equation}
and 
\begin{equation}
\label{3.42}
W_L \; \simeq \; t^2 \,  ( \ln L)^{1 - \frac{2}{3 a_1}} \;  
  \;  \; \;   \;  \;   \;  \;   \;  \;   triv. 
\end{equation}
Finally, for the specific  heat  one gets:
\begin{equation}
\label{3.43}
C_L \; \simeq \;  ( \ln L)^{ \frac{1}{3}} \;   
  \;  \; \;   \;  \;   \;  \;   \;  \;   \;  \;  \; \;  \;  \;   \;  \; pert. 
\end{equation}
and 
\begin{equation}
\label{3.44}
C_L \; \simeq \;  ( \ln L)^{1 - \frac{2}{3 a_1}} \;  
  \;  \; \;   \;  \;   \;  \;   \;  \;  \;  \;   \;  \;  triv. 
\end{equation}
In general, the critical behaviour of Ising-type systems can be analysed through its partition function zeroes.
The study of partition function zeroes  was initiated by Yang and Lee. The Lee-Yang theorem states that for ferromagnetic systems
all of the zeroes of the partition function in the external  magnetic field  lie on the imaginary axis for real temperatures. Fisher was the first
to analyse the zeroes in the complex temperature plane. Therefore, the partition function zeroes in the temperature plane are referred to  as Fisher zeroes,  while those in the complex plane of external fields as Lee-Yang zeroes. The finite-size scaling theory allow to determine the
logarithmic corrections to the zeroes of the partition function. The total free energy at the critical temperature in presence of an external
field is given by Eqs.~(\ref{3.37}) and (\ref{3.38}):
\begin{equation}
\label{3.45}
F_L(t=0,H) \; \simeq \;  c_2^{'}  \,  M^4 \, L^4 \,  ( \ln L)^{- 1} 
  \;  \; \;   \;  \;   \;  \;   \;  \;   \;  \;    \;  \;   \;  \;   pert. \; \& \; triv.
\end{equation}
where, from Eqs.~(\ref{3.35}) and (\ref{3.36}) at $t=0$ we have:
\begin{equation}
\label{3.46}
H \; \simeq \;  c_5 \,  M^3 \, ( \ln L)^{- 1}    \;  \; \;   \;  \;   \;  \;   \;  \; 
  \;   \;  \;   \;  \;    \;  \;   \;  \;     \;  \;    \;  \;   \;  \;   \;  \;   \;  \;    \;  \;   \;  \;     \;  \;    \;  \;   pert. \; \& \; triv.
\end{equation}
Whereupon we obtain:
\begin{equation}
\label{3.47}
F_L(t=0,H) \; \sim \;  H^{\frac{4}{3}} \, L^4 \,  ( \ln L)^{\frac{1}{3}} 
  \;  \; \;   \;  \;   \;  \;   \;  \;   \;  \;    \;  \;   \;  \;   pert. \; \& \; triv.
\end{equation}
The partition function is therefore:
\begin{equation}
\label{3.48}
Z_L(t=0,H) \; = \; Q( H^{\frac{4}{3}} \, L^4 \,  ( \ln L)^{\frac{1}{3}})
  \;  \; \;   \;  \;   \;  \;   \;  \;   \;  \;     pert. \; \& \; triv.
\end{equation}
From this last equation it follows that, if at some value of H the partition function vanishes, then:
\begin{equation}
\label{3.49}
 H^{\frac{4}{3}} \; = \;  \, L^{-4} \,  ( \ln L)^{- \frac{1}{3}}  \; Q^{-1}(0)
  \;  \; \;   \;  \;   \;  \;   \;  \;   \;  \;    \;  \;   \;  \;   pert. \; \& \; triv.
\end{equation}
or
\begin{equation}
\label{3.50}
 H_{j}  \; \sim  \;  \, L^{-3} \,  ( \ln L)^{- \frac{1}{4}} \;  \;  \;   \;  \;
   \;  \; \;   \;  \;   \;  \;   \;  \;   \;  \;    \;  \;   \;  \;     \;  \;   \;  \;   \;  \;    \;  \;    pert. \; \& \; triv.
\end{equation}
where $j$ indicates the index of the zero. For the Fisher zeroes, we note that from  Eqs.~(\ref{3.35}) and (\ref{3.36})  for $H=0$
one obtains for the total free energy:
\begin{equation}
\label{3.51}
F_L(t,H=0) \; \sim \;  L^{4} \, t^{2} \,  ( \ln L)^{\frac{1}{3}} 
  \;  \; \;   \;  \;   \;  \;   \;  \;   \;  \;    \;  \;   \;  \;  \;  \;  pert. 
\end{equation}
\begin{equation}
\label{3.52}
F_L(t,H=0) \; \sim \;  L^{4} \, t^{2} \,  ( \ln L)^{1 - \frac{2}{3 a_1}} 
  \;  \; \;   \;  \;   \;  \;   \;  \;   \;  \;    triv.
\end{equation}
Therefore, if  the partition function, which is the exponential of the free energy, vanishes, then for the jth zero:
\begin{equation}
\label{3.53}
  R_{j}^{-1}(0)  \; = \;  \, L^{4} \, t^2 \,   ( \ln L)^{\frac{1}{3}} \; 
  \;  \; \;   \;  \;   \;  \;   \;  \;   \;  \;    \;  \;   \;  \;   pert. 
\end{equation}
\begin{equation}
\label{3.54}
  R_{j}^{-1}(0)  \; = \;  \, L^{4} \, t^2 \,   ( \ln L)^{1 - \frac{2}{3 a_1}} \; 
  \;  \; \;   \;  \;   \;  \;   \;  \;   \;  \;    \;  \;    triv. 
\end{equation}
As a consequence we have:
\begin{equation}
\label{3.55}
 t_j  \; \sim \;  \, L^{-2} \,   ( \ln L)^{- \frac{1}{6}} \; 
  \;  \; \;   \;  \;   \;  \;   \;  \;   \;  \;    \;  \;   \;  \;  \;  \; \;   \;  \;   \;  \;   \;  \; pert. 
\end{equation}
\begin{equation}
\label{3.56}
  t_j  \; \sim \;  \, L^{-2} \,   ( \ln L)^{- \frac{1}{2} + \frac{1}{3 a_1}} \; 
  \;  \; \;   \;  \;   \;  \;   \;  \;   \;  \;    \;  \;   \;  \;   \;  \;  triv. 
\end{equation}
We now want to compare the finite-size analysis  to existing numerical calculations  in the literature for both asymptotic scaling laws. In the
broken phase we have only found numerical simulations dealing with the finite-size analysis for the zeroes of the partition function and
the specific heat. In the case of the Lee-Yang zeroes Ref.~\cite{Kenna:1994} reports the following measured exponent of
the logarithmic corrections:
\begin{equation}
\label{3.57}
\alpha_{meas}   \; = \;  -  \, 0.248(17)  \; \;  \; {\text{Lee-Yang \; zeroes}}
\end{equation}
For the Fischer zeroes Ref.~\cite{Kenna:1993} reports:
\begin{eqnarray}
\label{3.58}
\nonumber
\alpha_{meas}   \;  = \;  -  \, 0.217(12)  \; \;  8 \; \leq \, L \; \leq  24 \;  \; \; {\text{ Fisher zeroes}} 
\\
\alpha_{meas}   \;  = \;  -  \, 0.21(4)  \; \;  \, 12 \;   \leq \;  L \; \leq \;  24 \;  \; \;  {\text{Fisher zeroes}} 
\end{eqnarray}
In the case of the Lee-Yang zeroes both scaling laws give $\alpha_{theor} = - 1/4$ in excellent agreement with the measured exponent
Eq.~(\ref{3.57}). On the other hand, for the Fisher zeroes, using Eqs.~(\ref{3.55}),  (\ref{3.56}) and (\ref{3.10}) we have:
\begin{eqnarray}
\label{3.59}
\nonumber
\alpha_{theor}  \;  = \;  -  \; \frac{1}{6}   \;  \simeq \; - \; 0.167   \; \;   \; \; \; \;  \; \; \; \;  \; \;   \; \; \; \;  \; \;   \; \; \; \;  {\text{ Fisher zeroes}}  \; \; \; \;  \; \; \; \;  pert.
\\
\alpha_{theor}   \;  = \;  -  \; \frac{1}{2}  \; + \; \frac{1}{3 a_1} \; \simeq \; - \; 0.217(17)   \; \;   \; \; \; \;  \; \; {\text{ Fisher zeroes}}  \; \;   \; \; \; \;  \; \;   triv.
\end{eqnarray}
From this last equation we see that the triviality and spontaneous symmetry breaking scaling law is in excellent agreements with
the numerical outcomes, while the two-loop perturbative approach displays a sizeable deviation of at least two standard deviations. \\
Finally,  in Ref.~\cite{Lai:1990}  the four-dimensional Ising model is studied to probe the possibility 
of observing in Monte Carlo simulations the logarithmic corrections to the mean-field theory near criticality. In particular, 
by fitting   the specific heat $C_L$  as some power of $\ln L$, the  authors  of Ref.~\cite{Lai:1990}  reported:
\begin{equation}
\label{3.60}
\alpha_{meas}   \;  = \;    0.37(9)  \; \; \; \; \;   \; \; \;     6 \; \leq \, L \; \leq  14 \;  \; \; {\text{ specific heat}}  \; \; 
\end{equation}
On the other hand, from Eqs.~(\ref{3.43}) and (\ref{3.44}) we infer:
\begin{eqnarray}
\label{3.61}
\nonumber
\alpha_{theor}  \;  = \;  \frac{1}{3}   \;  \simeq \;  0.333   \; \;   \; \; \; \;  \; \; \; \;  \; \;   \; \; \; \; \; \;  \; \;  \; \;  \; \; \; \;  {\text{ specific heat}}  \; \; \; \;  \; \; \; \;  pert.
\\
\alpha_{theor}   \;  = \;   1 \; - \; \frac{2}{3 a_1}  \; \simeq \;  0.435(11)   \; \;   \; \; \; \;  \; \; \;   {\text{ specific heat}}    \; \; \; \;  \; \; \; \;  triv \;
\end{eqnarray}
From these  equations  we see that both  the triviality and spontaneous symmetry breaking  and the two-loop perturbative 
scaling laws are in  agreements with the numerical simulation results. \\
To summarise, we have shown that the triviality and spontaneous symmetry breaking picture compares remarkably well to several results
from non-perturbative numerical simulations. On the contrary, the two-loop renormalised perturbative approach suffers from serious
discrepancies with respect to numerical outcomes in the broken phase of the lattice implementation of the one-component
scalar field theory.
These results shed  doubts on the widely accepted opinion that the perturbative approach to the Higgs mechanism is fully supported 
by precise non-perturbative numerical simulations.
\section{The quantum liquid picture}
\label{s-4}
In the previous Sections we have discussed at length the one-component quantum scalar field with quartic self-interaction. Actually,
the scalar sector of the Standard Model involves four quantum scalar fields arranged in a weak doublet:
\begin{equation}
\label{4.1}
 \hat{\Phi}(x) \; = \;   
 \begin{pmatrix} \frac{\hat{\phi}_1(x) \, + \, i \, \hat{\phi}_2(x)}{\sqrt{2}}  \\  \frac{\hat{\phi}_3(x) \, + \, i \, \hat{\phi}_4(x)}{\sqrt{2}}  \end{pmatrix}  \;  \;  .
\end{equation}
Therefore the Lagrangian of the Standard Model scalar sector is assumed to be:
\begin{equation} 
\label{4.2} 
{\cal{L}}_{scalar} \;  = \;  \partial_\mu \hat{\Phi}^{\dagger} \;  \partial^\mu \hat{\Phi} \; - \;  \mu^2  \,   \hat{\Phi}^{\dagger}  \hat{\Phi}
\; - \;   \lambda_0  \, (\hat{\Phi}^{\dagger}  \hat{\Phi})^2
\end{equation}
with a tachyonic mass term $\mu^2 < 0$ and a positive quartic self-coupling $\lambda_0 > 0$. Evidently the Lagrangian Eq.~(\ref{4.2})
corresponds to a $O(4)$-symmetric  self-interacting scalar field theory. As compared to the one-component model, the solution of the 
$O(n)$-symmetric model is not much more difficult. In the broken symmetry phase, the presence of (massless) Goldstone bosons
 for $n > 2$ gives rise to a number of complications. Nevertheless, it has been well established by numerical simulations the triviality
of  the $O(4)$ symmetric model  in the broken phase (see Ref.~\cite{Luscher:1989} and references therein).  
The emerging physical picture of the $O(4)$ model thus obtained is practically the same as for the one-component model treated earlier.
In particular, as in the one-component model, the triviality of the theory does not allow us to implement the spontaneous
symmetry breaking mechanism and, as a result of the triviality, we should end with four massless scalar fields.  Fortunately the
scalar fields of the Standard Model are also coupled to the gauge vector bosons of the electroweak gauge group $SU(2) \otimes U(1)$.
This is easily accomplished by replacing in Eq.~(\ref{4.2}) the derivatives $\partial_\mu$ with the covariant derivatives $D_\mu$.
However, in the absence of gauge fixing, a local gauge symmetry cannot break spontaneously,  as we know from Elitzur's  
theorem~\cite{Elitzur:1975}. As a consequence, to enforce the symmetry breaking of the gauge symmetries we are forced to
make a gauge choice. Since the physics cannot depend on the gauge fixing, we are free do adopt the physical gauge where
three scalar filed are gauged away (unitary gauge). Aa a consequence the quantum scalar fields Eq.~(\ref{4.1}) can be written as:
\begin{equation}
\label{4.3}
 \hat{\Phi}(x) \; = \;   
 \begin{pmatrix} 0  \\  \frac{\hat{\phi}(x) }{\sqrt{2}}  \end{pmatrix}  
\end{equation}
where $\hat{\phi}(x)$ is a one-component quantum scalar field. We see, thus, that the Lagrangian Eq.~(\ref{4.2}) reduces
to the Lagrangian in Eq.~(\ref{2.1}). As discussed in Sect.~\ref{s-2}, to trigger the spontaneous symmetry breaking we need some amount
 of non-Gaussian fluctuations in the scalar quantum vacuum. Again we face with the problem that, within the pure scalar sector, we do not have a physical mechanism to generate a positive self-coupling needed to trigger the Bose-Einstein condensation of the quantum scalar field.
 Actually, in Ref.~\cite{Cea:2020} we showed that  the interactions of the scalar field $\hat{\phi}$ with vector bosons and fermion fields
 of the standard Model  will induce an effective positive scalar self-coupling  $\lambda_{eff}$ whose scale evolution is governed 
by the running coupling constants $g(\mu) , g'(\mu), \lambda_f(\mu)$,  where $g$ is the $SU(2)$ gauge coupling, $g'$ the $U(1)$ gauge
coupling and  $\lambda_f(\mu)$ indicates the generic coupling of $\hat{\phi}$ to the fermion fields. We are led, then, to the following
effective scalar Lagrangian:
\begin{equation} 
\label{4.4} 
{\cal{L}}\;  = \; \frac{1}{2} \, (\partial_\mu \hat{\phi})^2 \;- \;  \frac{1}{4} \,  \lambda_{eff}  \,  \hat{\phi}^4  \; .
\end{equation}
We have already seen that the presence of a positive quartic self-coupling, irrespective to the actual value
of the self-coupling,  leads  to the Bose-Einstein condensation of the
scalar field $\hat{\phi}$:
\begin{equation} 
\label{4.5} 
\hat{\phi}(x)  \; = \;   \hat{H}(x)   \;  + \; v
\end{equation}
characterised  by a finite scalar condensate $v$ and a massive quantum scalar field $H(x)$ (the Higgs field) with mass finitely related
to the condensate strength. The strength of the condensate is an arbitrary parameter of the theory. Once identified $v$ with the Standard Model
condensate strength, $v \simeq 246$ GeV, the mass of the fluctuating Higgs field has been estimated by precise non-perturbative
numerical simulation, Eq.~(\ref{3.22}). As a matter of fact, a more precise value of the Higgs field has been obtained in
Ref.~\cite{Cea:2019} where, for the first time, it was enlightened in the preliminary LHC Run 2 data
some evidences of  a rather broad resonance structure around  700 GeV consistent with a heavy Higgs boson. 
According to Ref.~\cite{Cea:2019}, in the following  we shall assume:
\begin{equation} 
\label{4.6} 
m_H \;  \simeq \; 730  \; \; \text{GeV} \; 
\end{equation}
that, by the way, is fully consistent with the lattice determination, Eq.~(\ref{3.22}). \\
We have reached  the remarkable result that the massless quantum scalar fields of the Standard Model, once coupled to the
gauge vector bosons and fermions, undergoes dynamically the Bose-Einstein condensation needed to implement the Higgs
mechanism.  
Note that  in our approach we are not invoking the unphysical Mexican-hat classical potential. The symmetry breaking
mechanism is generated by non-Gaussian  fluctuations in the scalar quantum vacuum generated by a positive quartic self-coupling.
Since our mechanism does not depend on the actual value of the positive quartic self-coupling, we see that there is no stability
problem for the vacuum condensation. On the other hand, it should be clear that  our  $\lambda_{eff}$  is not washed out by the
scalar quantum fluctuations since  the scale evolution is dictated by the renormalisation group evolution of the gauge and Yukawa couplings.
Actually, it is known that the scale evolution of $\lambda_{eff}$  dictates that the quartic self-coupling becomes negative at some
high-energy scale $\Lambda_{eff} \sim 10^{10}$ GeV. This means that our mechanism can be realised only if  $v \lesssim \Lambda_{eff}$,
condition amply satisfied in the Standard Model. \\
\begin{figure}
\vspace{-0.5cm}
\begin{center}
\includegraphics[width=0.8\textwidth,clip]{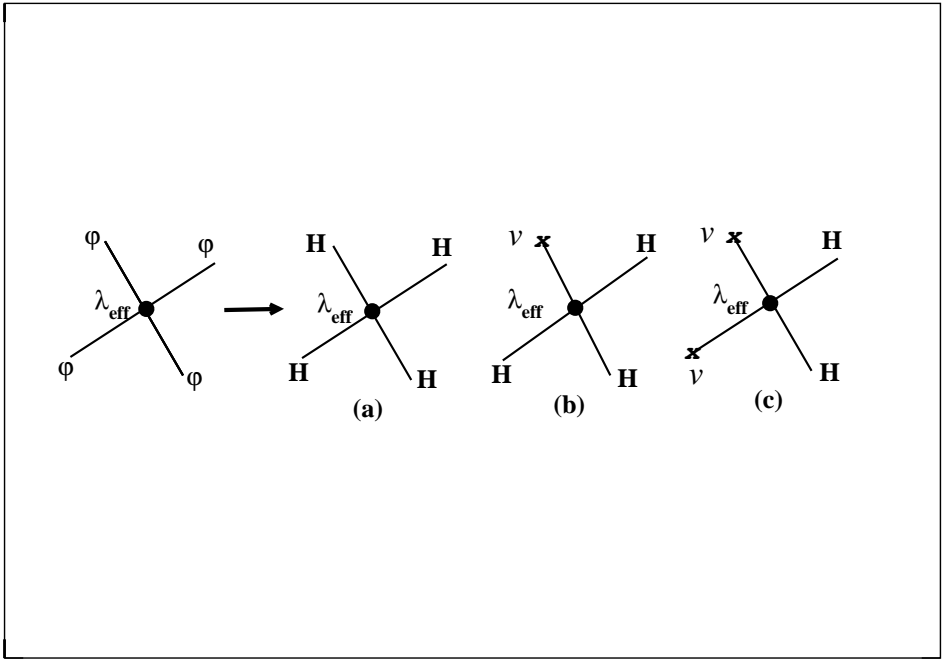}
\end{center}
\vspace{-0.5cm}
\caption{\label{quartic} 
Self-couplings of the  H scalar field.}
\end{figure}
Up to now we have found that the triviality and spontaneous symmetry breaking picture leads in the Standard Model to the 
scalar Bose-Einstein condensation giving rise to an almost free massive scalar fluctuating Higgs field $\hat{H}$ with a rather
large mass given by Eq.~(\ref{4.6}). This picture is modified substantially  by the presence of a small positive quartic
self-coupling~\cite{Cea:2020}. In fact, the quartic self-coupling, after taking into account Eq.~(\ref{4.5}), leads to new interacting 
terms for the Higgs field. In Fig.~\ref{quartic} we show how the self-coupling generates new interaction terms for the Higgs field.
Actually in   Fig.~\ref{quartic}  we are not displaying the interaction term linear in $\hat{H}$. Indeed, since the fluctuating quantum field
$\hat{H}$ involves only modes with $\vec{k} \neq 0$, this term vanishes.  There is a further term proportional to $v^4$ that merely
adds a contribution to the vacuum energy density that, now, is given by:
\begin{equation} 
\label{4.7} 
{\cal{E}}  \; \simeq  \;  - \;  \frac{m_H^4}{128 \pi^2 } \;  +  \;  \frac{\lambda_{eff}}{4} \; v^4 \;. 
\end{equation} 
In Ref.~{\cite{Cea:2020} we argued that $\lambda_{eff} \lesssim  \lambda_{SM}$, where  $\lambda_{SM}$ is the quartic self-coupling
in the standard perturbative approach with the Mexican-hat classical potential.  So that the vacuum energy density in Eq.~(\ref{4.7})
is still negative. \\
Looking at Fig.~\ref{quartic} we see that the Higgs field acquires quartic and cubic self-interaction terms.
There is a further  interaction term quadratic in the Higgs field.  This quadratic term implies that the Higgs field may
undergo  interactions with the vacuum condensate that generate a new coherent propagating condensate
excitations with mass:
\begin{equation}
\label{4.8}
m^2_h   \; \sim  \; \lambda_{eff}  \; v^2 \;  \sim \; 10^2 \; \text{GeV}    \; .
\end{equation}
In Ref.~\cite{Cea:2020} this  condensate excitation was  identified with the new LHC  scalar resonance  with mass 125 GeV. 
This lead us to look at the Higgs condensate  as a quantum liquid analogous to the Bose-Einstein condensate in superfluid He II. 
Indeed,  the low-lying excitations of the Higgs condensate resembled two Higgs bosons that  correspond to the relativistic version of
 the phonons and rotons in superfluid He II. In the dilute gas approximation, that is 
the relevant regime for the LHC physics, these   excitations of the Higgs condensate behave as  two  Standard Model Higgs bosons
denoted with h and H respectively. The light Higgs boson has a mass given by Eq.~(\ref{4.8}), while the mass of the heavy Higgs boson
is given by  Eq.~(\ref{4.6}).
 Note that the mass of the h Higgs boson is naturally related to the weak scale $ v \sim 10^2$ GeV, i.e.  there is no  naturalness problem,
 namely we do not need the fine tuning of  any parameter to obtain a mass of order $10^2$ GeV. \\
We would like to emphasise  that we are not saying that there are two different elementary quantum Higgs fields.
On the contrary, we have a unique local quantum Higgs field. However, since the scalar condensate
behaves like the He II quantum liquid, when the Higgs field acts on the condensate
it can give rise to two elementary excitations, namely the phonon-like and roton-like excitations
corresponding to long-range collective and localised disturbances of the condensate, respectively.
These elementary condensate excitations behave as weakly interacting scalar
fields with vastly different mass.  These two Higgs bosons will interact with the gauge vector bosons.
We already pointed out~\cite{Cea:2020} that the couplings of the Higgs condensate elementary excitations to the
gauge vector bosons are fixed by the gauge symmetries. As a consequence, both the Higgs bosons h and H  will be
coupled to gauge bosons as in the usual perturbative approximation of the Standard Model. \\
As concern the coupling to fermion fields, if we admit the presence of the Yukawa terms in the Lagrangian, then we are led to an effective
Yukawa lagrangian:
\begin{equation}
\label{4.9}
{\cal L}_{Y}^{eff}(x) = 
\sqrt{Z^h_{wf}}  \; \frac{\lambda_f}{\sqrt{2}} \;  \hat{\bar{\psi}}_f(x)  \,   \hat{\psi}_f(x) \;   \hat{h}(x)   +  
\sqrt{Z^H_{wf}} \;  \frac{\lambda_f}{\sqrt{2}}  \; \hat{\bar{\psi}}_f(x)  \,  \hat{\psi}_f(x) \;  \hat{H}(x)  \; , 
\end{equation}
where  $\hat{\psi}_f(x)$ indicates  a generic fermion quantum field and  the Yukawa coupling satisfies the usual relation:
\begin{equation}
\label{4.10}
\lambda_f \; = \;  \frac{\sqrt{2} \, m_f }{v} \; .
\end{equation}
In  Eq.~(\ref{4.9})  $Z^h_{wf}$ and   $Z^H_{wf}$ are wavefunction renormalisation constant that, roughly,  take care
of the eventual mismatch in the overlap between the fermion and quasiparticle wavefunctions. 
In Ref.~\cite{Cea:2020}  we  were able to fix  these constants from a  comparison with the experimental observations.  As a  result we argued  
that:
\begin{equation}
\label{4.11}
Z^h_{wf}  \; \simeq \; 1 \;  \; \; , \; \; \;  Z^H_{wf} \; \simeq  \; \frac{m_h}{m_H} \; \simeq  0.171 \; .
\end{equation}
A notable consequence of   Eq.~(\ref{4.11}) is  that our light Higgs boson h is practically indistinguishable from the perturbative Higgs boson.  
\section{Comparison with experimental observations}
\label{s-5}
Once established that  the perturbations of the scalar condensate due to the quantum Higgs field behave as two independent
weakly interacting massive scalar fields, we need to investigate  the experimental signatures  and the interactions of these
Higgs  condensate elementary excitations.  Obviously, the most striking consequence of our approach is the prevision  of an additional 
 heavy  Higgs boson~\footnote{Interestingly enough, in Ref.~\cite{Romatschke:2023} it has been enlighten the presence of a bound
 state with a mass of approximately 1.84 M $\sim $ 230  GeV.}.  
As we already said, the light Higgs boson is the
natural candidate for the new LHC scalar resonance at 125 GeV. On the other hand, our previous phenomenological analysis of the
preliminary LHC Run 2 data in the golden channel~\cite{Cea:2019} hinted at  the presence of a broad scalar resonance with
central mass of 730 GeV. 
The aim of this Section is to present the comparison of our theoretical proposal to the experimental observations from both
the ATLAS and CMS LHC Collaborations based on the full Run 2 datasets. Considering that our light Higgs boson h is practically
indistinguishable from the perturbative Higgs boson, we shall restrict ourself to the heavy Higgs boson H. \\
The phenomenological signatures of a massive  Higgs boson are determined by the couplings with the gauge and fermion fields of the
Standard Model.  As already argued in Sect.~\ref{s-4}, the couplings of the Higgs  field to the gauge vector bosons 
are dictated by the gauge symmetries, so that  the couplings of the H Higgs boson to the gauge vector bosons are the same as in perturbation theory 
 notwithstanding the non-perturbative scalar condensation driving the spontaneous breaking of the gauge symmetries. The unique difference
arises from the Yukawa coupling to the fermions. Given the rather large mass of the H Higgs boson, the only relevant Yukawa coupling is the
coupling to the top quark. According to Eq.~(\ref{4.9}) the top Yukawa coupling of the H Higgs boson is suppressed  by a factor
$\sqrt{Z^H_{wf}}$ with respect to perturbation theory.  \\
The Higgs boson can be produced in many different ways. In general,  there are four common ways to produce the Higgs boson in proton-proton colliders: gluon-gluon fusion, vector-boson fusion, associated production with a vector boson and associated production with a pair of top quarks.
However, the main production mechanisms are  by  gluon-gluon fusion and vector-boson fusion. Therefore, in the following we
shall restrict ourself to these production mechanisms.  
To evaluate the Higgs event production at LHC we need the inclusive Higgs production cross section.
Observing that  the couplings of the H boson to vector bosons are the same as those of a Standard Model Higgs boson, the H boson
 production cross section by vector-boson fusion is the same as in the perturbative Standard Model calculations. On the other hand,
for  the gluon-gluon fusion production mechanism,  considering that   the gluon coupling to the Higgs boson in the Standard Model is 
mediated mainly  by triangular loops of top and bottom quarks, it follows that the inclusive production cross section is given by the Standard Model
cross section suppressed by a factor  $Z^H_{wf}$.  Therefore, within our approximations, the total inclusive cross section 
for the production of the H Higgs boson we can  write:
\begin{equation}
\label{5.1}
\sigma(p \; p \;  \rightarrow \; H) \; \simeq \;    \sigma_{VBF}(p \; p \;  \rightarrow \; H)
\; + \;  \sigma_{GGF}(p \; p \;  \rightarrow \; H)  \; ,
\end{equation}
where  $\sigma_{VBF}$ and   $\sigma_{GGF}$ are the vector-boson fusion and gluon-gluon fusion inclusive cross
sections, respectively. 
In the Standard Model the calculations of the cross sections computed at next-to-next-to-leading 
and next-to-leading order  for  a high mass Higgs boson with Standard Model-like couplings
 at $\sqrt{s} = $ 13  TeV are provided by the LHC Higgs Cross Section Working Group~\cite{Hcross13Tev}.
\begin{figure}
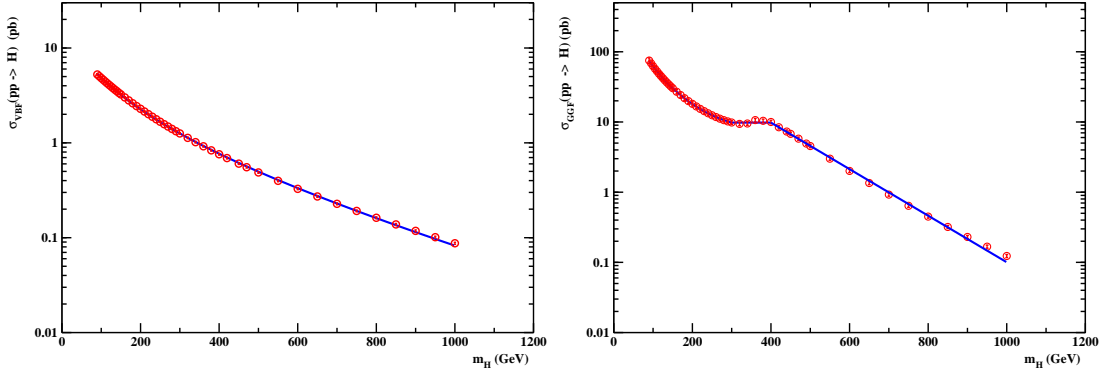

\vspace{-0.5cm}
\begin{center}
\includegraphics[width=0.45\textwidth,clip]{Fig5a.eps}
\includegraphics[width=0.45\textwidth,clip]{Fig5b.eps} 
\end{center}
\caption{\label{Fig5} 
(Color online) Inclusive H Higgs boson  production cross sections  for the VBF  processes (left panel) and
GGF processes (right panel) at $\sqrt{s} = $ 13  TeV.  The data points  have been taken from Ref.~\cite{Hcross13Tev}.
The (blue)  continuous lines correspond to  Eqs.~(\ref{5.2}) and (\ref{5.4}) respectively.}
\end{figure}
As concern the Standard Model gluon-gluon fusion cross section  we found~\cite{Cea:2019} that  this cross section can be
 usefully interpolated by (see Fig.~\ref{Fig5}, right panel):
\begin{equation}
\label{5.2}
 \sigma_{GGF}^{SM}(p \; p  \rightarrow H)  \; \simeq \;  
 \left\{ \begin{array}{ll}
 \;  \left (  \frac{ a_1}{ m_H} 
 \; + \; a_2 \; m_{H}^3  \right )  \;  \exp (-  a_3 m_{H})  \; \; &  m_{H}  \; \leq \; 300  \; GeV 
  \\
 \; \; \;  \; a_4  \;  & 300 \; GeV    \leq  m_{H}   \leq  400 \; GeV
  \\
 \; \;  \; \;a_4 \;  \exp \big [ - a_5 ( m_{H} - 400 \; GeV) \big ]  \; \; &  400  \; GeV \; \leq \; m_{H}
\end{array}
    \right.
\end{equation}
where $m_{H} $  is expressed in  GeV and:
\begin{eqnarray}
\label{5.3}
a_1 \simeq 1.24 \, 10^4 \; pb \; GeV \;  \; , \; \;  a_2 \simeq 1.49 \, 10^{-6} \; pb \; GeV^{-3} \; , \;  
\nonumber \\
a_3 \simeq 7.06 \, 10^{-3} \;  GeV^{-1} \; , \;  \; a_4 \simeq 9.80 \;\, pb  \; , \hspace{2.2 cm}
\\ \nonumber
a_5 \simeq 7.63 \, 10^{-3}  \; GeV^{-1}  \; . \hspace{5.05 cm}
\end{eqnarray}
Likewise   the Standard Model vector-boson fusion cross section  can be parametrised as  (see Fig.~\ref{Fig5}, left  panel):
\begin{equation}
\label{5.4}
 \sigma_{VBF}^{SM}(p \; p  \rightarrow  H) \; \simeq \;    \bigg ( b_1 \; + \;  \frac{ b_2}{ m_{H}} 
 \; + \; \frac{b_3}{ m_{H}^2}  \bigg )  \;  \exp (-  b_4 \;  m_{H} )   \; ,
\end{equation}
with:
\begin{eqnarray}
\label{5.5}
b_1 \simeq - 2.69 \, 10^{-6}  \; pb   \;  \; , \; \;  b_2 \simeq 8.08 \, 10^{2} \; pb \; GeV \; , \hspace{1.15 cm}
 \nonumber \\
b_3 \simeq - 1.98 \, 10^{4}  \; pb \;  GeV^{2}  \; \;  , \;  b_4 \simeq  2.26 \, 10^{-3} \; GeV^{-1} \; .  \; \; \,
\end{eqnarray}
\begin{figure}
\vspace{-0.5cm}
\begin{center}
\includegraphics[width=0.8\textwidth,clip]{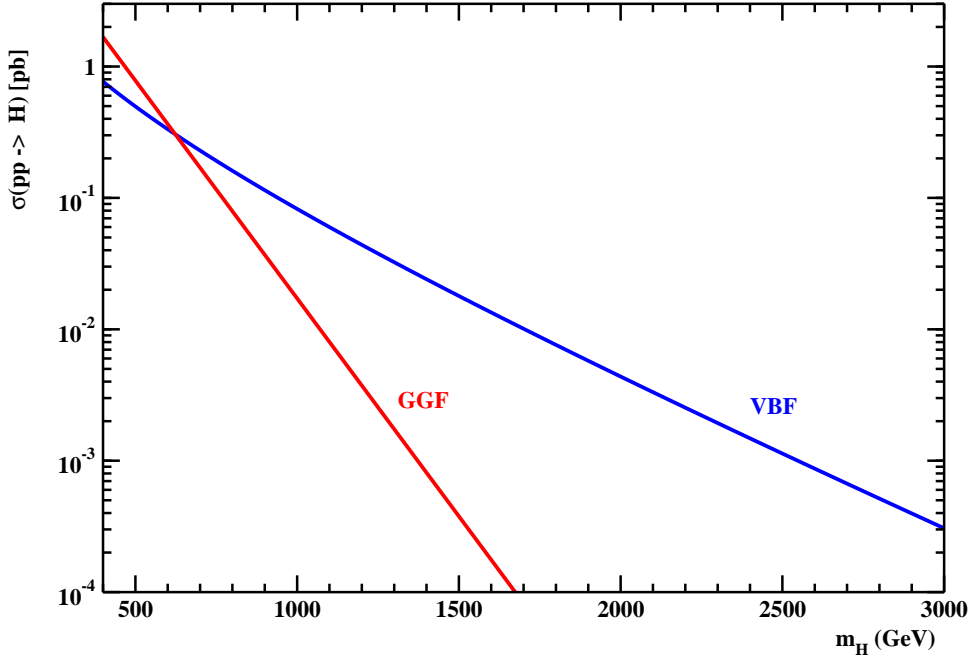}
\end{center}
\caption{\label{Fig6} 
(Color online) Inclusive H Higgs boson  production cross sections  for the VBF  processes (blue continuous line) and
GGF processes (red continuous line) as a function of $m_H$ at $\sqrt{s} = $ 13  TeV.}
\end{figure}
We would like to stress that in theoretical calculations of the production cross section, in the mass range relevant to us, 
there is a typical $ \pm 5  \, \% $ uncertainty  due to the choice of the parton distributions,  the QCD scale and 
the strong interaction coupling. In the following, usually, we will  not  take into account these theoretical uncertainties. \\
Our previous discussion lead us to assume that to a  good approximation we can write:
\begin{equation}
\label{5.6}
  \sigma_{VBF}(p  p   \rightarrow  H)  \simeq   \sigma_{VBF}^{SM}(p p   \rightarrow  H) \;   \; , \; \; 
  \sigma_{GGF}(p p  \rightarrow  H)  \simeq   Z^H_{wf} \;  \sigma_{GGF}^{SM}(p p  \rightarrow  H) \; .
\end{equation}
In Fig.~\ref{Fig6} we compare the VBF  and GGF production cross sections given by Eq.~(\ref{5.6}) after taking into
account the value of $Z^H_{wf}$ in Eq.~(\ref{4.11}). It is evident that the main production mechanism
of the heavy H Higgs boson is  by the VBF  processes since at $m_H  \; \simeq \; 730$  GeV it results:
\begin{equation}
\label{5.7}
  \sigma_{VBF}(p  p   \rightarrow  H) \;  \simeq \;   2 \;   \sigma_{GGF}(p p  \rightarrow  H)  \;   \; .
\end{equation}
In order to determine the phenomenological signatures of the massive H Higgs boson we need to examine the decay modes.
Given the rather large mass of the heavy Higgs boson, the main decay modes are the decays into two massive
vector bosons (see, e.g., Refs.~\cite{Gunion:1990,Djouadi:2008}):
\begin{equation}
\label{5.8}
\Gamma( H \; \rightarrow \; W^+ \, W^-)  \; \simeq  \;  \frac{G_F \, m^3_{H}}{8 \pi \sqrt{2}} \;
 \sqrt{1 - \frac{4 m^2_W}{m^2_{H}}} \;  \bigg ( 1 - 4 \,  \frac{m^2_W}{m^2_{H}} + 12 \, \frac{ m^4_W}{m^4_{H}}
 \bigg ) \; 
\end{equation}
and
\begin{equation}
\label{5.9}
\Gamma( H  \;  \rightarrow \;  Z^0 \, Z^0) \; \simeq \;   \frac{G_F \, m^3_{H}}{16 \pi \sqrt{2}} \;
 \sqrt{1 - \frac{4 m^2_Z}{m^2_{H}}} \; \bigg ( 1 - 4 \,  \frac{m^2_Z}{m^2_{H}} + 12 \, \frac{ m^4_Z}{m^4_{H}}
 \bigg )  \; . 
\end{equation}
The couplings of the H Higgs boson to the fermions are given by the Yukawa couplings. As we said before, for heavy Higgs 
the only relevant fermion coupling is  the top Yukawa coupling. The width for the decays of the H boson into a $t \bar{t}$ pairs is easily found~\cite{Gunion:1990,Djouadi:2008}:
\begin{equation}
\label{5.10}
\Gamma( H \rightarrow \; t \, \bar{t}) \; \simeq \; Z^H_{wf}  \;  \frac{3 \, G_F \, m_{H} m^2_t}{4 \pi \, \sqrt{2}} \;
\bigg ( 1 - 4 \,  \frac{m^2_t}{m^2_{H}}  \bigg )^{\frac{3}{2}}  \; .
\end{equation}
Obviously, there are several other   decay modes that, however, are expected to be negligible. 
So that, to a good approximation, the heavy Higgs  boson total width is given by:
\begin{equation}
\label{5.11}
\Gamma_{H}  \; \simeq \; \Gamma( H \rightarrow W^+ \, W^-)  \; + \; \Gamma( H \rightarrow Z^0 \, Z^0)  \; + \;
 \Gamma( H \rightarrow t \, \bar{t})  \; .
\end{equation}
We see that  almost all the  decay modes  are given by the decays into W$^+$W$^-$ and Z$^0$Z$^0$. Moreover we have:
\begin{equation}
\label{5.12}
Br(H \rightarrow W^+ W^-) \; \simeq \; 2 \; Br(H \rightarrow Z^0 Z^0) \; .
\end{equation}
The above discussion   show that  our heavy Higgs boson is a rather broad resonance  with a total width $\Gamma_H$ that is about
30 \% of its mass. This poses challenging theoretical problem in dealing with a resonance with such a huge width. A convenient way is 
to use the Breit-Wigner formula~\cite{Breit:1936} to model the resonance  in terms of the physical mass and width.
More precisely,  in the resonance rest frame we may  estimate the probability to find the H resonance in the energy interval 
$(E, E + dE)$ by means of the modified Lorentzian distribution: 
\begin{equation}
\label{5.13}
L_ {H} (E) \;  \simeq  \;  \; \frac{1}{ 1.0325 \;  \pi} \; \frac{\frac{\Gamma_{H}(E)}{2}}{\big (E \; - \; 730 \; GeV \big )^2 \; + 
\;  \big ( \frac{\Gamma_{H}(E)}{2} \big )^2} \;  \;   \;  \; , 
\end{equation}
where $\Gamma_{H} ( E)$ is given by Eq.~(\ref{5.11}), and the normalisation is  such that:
\begin{equation}
\label{5.14}
\int^{\infty}_{0} \; L_ {H} (E)  \;  dE \; \; = \; \; 1 \; \; .
\end{equation}
Note that, in the limit $\Gamma_H \rightarrow 0$,   $L_H(E)$ reduces to $\delta(E - 730 \,  GeV)$.   \\
In general, we are interested in collider processes $ p p \rightarrow H \rightarrow X$. In particular, we wish to estimate  the invariant mass 
spectrum of the particles produced in the Higgs decays. Let  $N_{H} (E_{1},E_{2} )$ be the number of events produced by the H Higgs decays
in the energy interval $(E_1, E_2)$.  Then, we   have:
\begin{equation}
\label{5.15}
 N_{H} (E_{1},E_{2} )  \; \simeq \; {\cal{L}} \;  \int^{E_2}_{E_1} \;
  \sigma(p \; p \;   \rightarrow \; H )   Br(H \rightarrow X) \;  \varepsilon(E) \;  \; L_ {H} (E)  \;  dE  
 \;  \; , 
\end{equation}
where  ${\cal{L}}$ is the   integrated luminosity. The parameter $ \varepsilon(E)$  accounts for  the efficiency 
of trigger, acceptance of the detectors, the kinematic selections, and so on.   In general,
 $ \varepsilon(E)$ is expected to depend on the energy, the selected channel and the detector.  The above approximations should
 furnish a reasonable estimate of the invariant mass distribution for values of the invariant mass $M$  in a range $\Delta M \sim 10^2$ GeV
 around $m_H$.
\subsection{Comparison with the ATLAS Run 2 datasets}
\label{s-5-1}
\begin{figure}
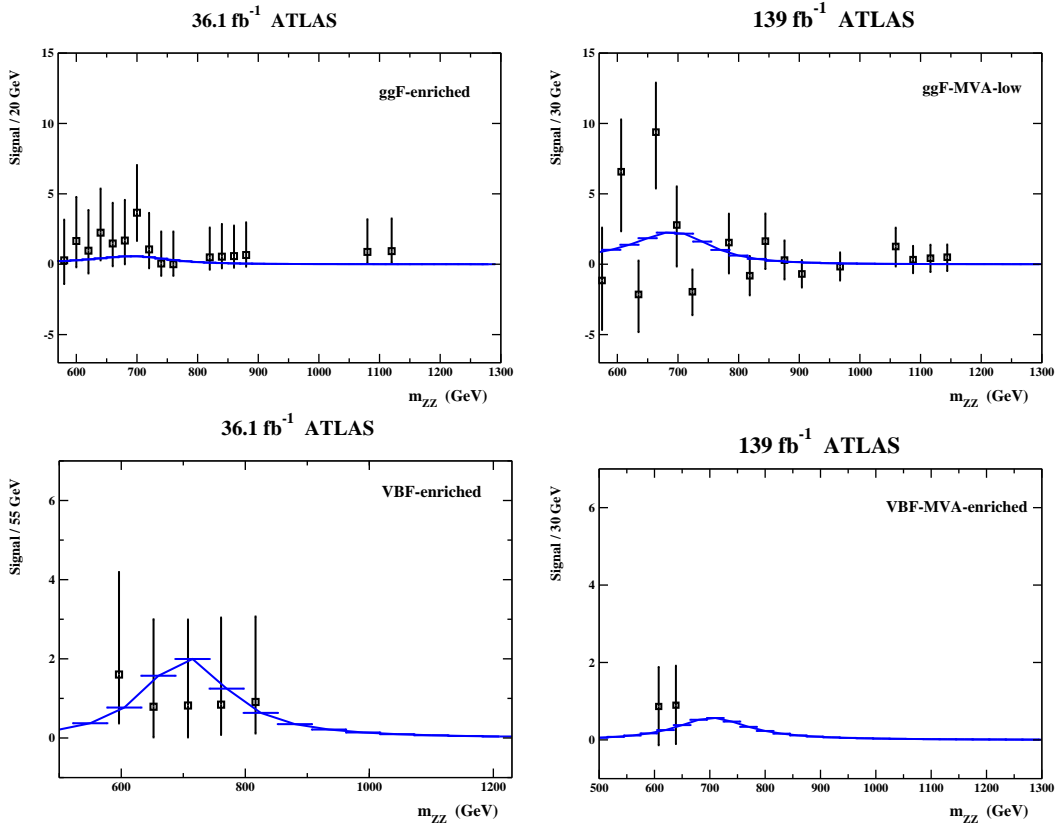

\begin{center}
\vspace{-1.5cm}
\includegraphics[width=0.42\textwidth,clip]{Fig7a.eps}
\hspace{0.2 cm}
\includegraphics[width=0.42\textwidth,clip]{Fig7b.eps} 
\vspace{0.1cm}
\includegraphics[width=0.42\textwidth,clip]{Fig7c.eps}
\hspace{0.2 cm}
\includegraphics[width=0.42\textwidth,clip]{Fig7d.eps} 
\end{center}
\vspace{-0.7cm}
\caption{\label{Fig7} 
(Color online) Signal distributions   versus  the invariant mass $m_{ZZ}$ in the high-mass region
 $m_{ZZ} \gtrsim 600$ GeV   for the GGF and VBF processes with integrated luminosity $36.1 fb^{-1}$
 and $139  fb^{-1}$.   The (blue) continuous lines  are the expected theoretical distributions for the H Higgs boson.}
\end{figure}
In the present Section we intend to contrast our theoretical proposal of an additional Standard Model heavy Higgs boson 
to the experimental outcomes from the LHC ATLAS Collaboration. Actually, in a recent paper~\cite{Cea:2021}
we attempted  a comparison of our proposal to the ATLAS full Run 2 datasets in the main decay channels
$H \rightarrow WW, ZZ$.   As concern the decays  $H \rightarrow ZZ$   in the golden channel here we merely restrict ourself to a summary of 
Ref.~\cite{Cea:2021}.  To this end, let us consider the golden channel  $H  \rightarrow  ZZ  \rightarrow  \ell^+ \ell^- \ell^+ \ell^-$, 
$\ell$ being an electron or a muon. 
It is well established that  the golden channel has very low branching ratio,
nevertheless  the presence of leptons allows to efficiently reduce the huge background mainly due  to diboson production.
Indeed, the four-lepton channel, albeit rare, has the clearest and cleanest signature of all the
possible Higgs boson decay modes due to the small background contamination. 
Since it is assumed that an additional Higgs boson would be produced
predominantly via gluon-gluon fusion (GGF) and vector-boson fusion (VBF) processes,  the
events are classified into GGF and VBF categories and results are interpreted separately for the GGF and VBF production modes.
A search for a new high-mass resonance decaying into electron or muon pairs has been performed 
the ATLAS  experiment  using data collected at $\sqrt{s} = 13$  TeV  corresponding to an integrated luminosity of 
36 fb$^{-1}$~\cite{Aaboud:2018} and upgraded to 139 fb$^{-1}$~\cite{Aad:2021}.
In order to be sensitive to the VBF production mode, for the 36.1 $fb^{-1}$ data set, 
the ATLAS Collaboration~\cite{Aaboud:2018} classified the events into four categories, namely  one for the
VBF production mode and three for the GGF production mode.
 If an event has two or more jets with $p_T$ greater than 30 GeV, with the two leading jets $j_1,j_2$ well separated in
the pseudo-rapidity $\eta$,  i.e. $| \eta_{j_1j_2} |  > 3.3 $,  and having an invariant mass $m_{j_1j_1}  > 400$  GeV, 
then this event is classified into the VBF-enriched category.
 Otherwise the event is classified into one of the GGF-enriched categories. Such classification is
used only in the search for a heavy scalar produced within the narrow width approximation.
On the other hand, for the full Run 2 data set the event classification targeting different production processes has been optimised
using machine learning algorithms~\cite{Aad:2021}.
 More precisely, to improve the sensitivity in the search of a heavy Higgs signal produced either in the VBF 
or in the GGF production mode, it was used two classifiers,  a  VBF classifier  and a GGF classifier. 
These classifiers were built with deep neural networks. The networks were trained  by means of  simulated signal events
from a heavy Higgs boson with masses ranging from 200 GeV up to 1400 GeV in the narrow width approximation,
 and from the Standard Model continuous  ZZ background. \\
In Fig.~\ref{Fig7}, adapted from Figs.~2 and 5 of Ref.~\cite{Cea:2021},  we show the signal invariant-mass distribution for 
the golden channel for the  GGF category corresponding to 36 fb$^{-1}$  (upper left panel) and 139 fb$^{-1}$ (upper right panel) 
and VBF category  ( 36 fb$^{-1}$ bottom left panel,  139 fb$^{-1}$ bottom right panel)  in the high invariant mass
region $m_{ZZ} \gtrsim 600$ GeV.
The signal distributions can be easily obtained  by subtracting  from  the event distributions  the  irreducible  ZZ  backgrounds
that constitute the main source of the Standard Model background in the high  invariant-mass region. 
We, also, compare in   Fig.~\ref{Fig7} the observed signal distributions to our theoretical proposal obtained by means of
Eq.~(\ref{5.15}). 
 For the 2016 ATLAS data set, we see that the expected signal histogram is perfectly compatible with the data, 
 but it is evident  that the integrated luminosity is too low to claim an evidence of our heavy Higgs boson.
As concern the full data set,  for the comparison we have taken into account that the selection cuts applied by the ATLAS Collaboration to
the full Run 2 data set are more stringent with respect to the preliminary data.  As a consequence, the resulting efficiency  to detect
a high-mass broad resonance is drastically reduced.
This lead us to argue  that there was not enough sensitivity to detect the expected signal at least in this channel
 (this last point is thoroughly discussed in  Ref.~\cite{Cea:2021}). \\
According to Eq.~(\ref{5.12})  the main decay mode of our heavy Higgs boson is the decays into two W vector bosons. Thus,
the most stringent constraints should arise from the experimental searches for a heavy Higgs boson
decaying into two W gauge bosons. In fact, in our previous paper~\cite{Cea:2020} we compared our theoretical expectations to
the   search for neutral heavy resonances  in the $WW  \rightarrow   e \nu \mu \nu$    decay channel  performed by the ATLAS Collaboration
using proton-proton  collision data at $\sqrt{s} = 13$ TeV and  corresponding to an integrated luminosity of 36.1  
$fb^{-1}$~\cite{Aaboud:2018}.
In Ref.~\cite{Cea:2021} we  analysed  the full Run 2 dataset  presented by the ATLAS Collaboration~\cite{Aad:2020}
on the search of  heavy resonances decaying into two vector bosons WW,  ZZ or WZ using collected data corresponding
 to  an integrated luminosity of 139  $fb^{-1}$. Actually, in Ref.~\cite{Aad:2020} the research focused 
on  a  heavy neutral scalar resonance, called the Radion,   which appears in some
theoretical  models and which, indeed,  can decay into WW or ZZ  with a branching ratio approximatively given by  Eq.~(\ref{5.12}). 
 Moreover, the  Radion-like scalar resonances couple to the Standard Model fermions  and gauge bosons with  strengths similarly to a heavy Higgs
boson. Considering that these heavy scalar resonances  have  a rather  narrow widths, in our previous paper~\cite{Cea:2021}
we considered  the observed limits on the production processes  presented in Ref.~\cite{Aad:2020}
as indicative of the production of a heavy Standard Model heavy Higgs boson in the narrow width approximation. 
Since recently the ATLAS Collaboration has presented the results of a search for heavy neutral Higgs-like resonances decaying into two W 
 bosons, which subsequently decay into the $e \nu \mu \nu$  final state~\cite{ATLAS:2022},  it is worthwhile here to compare 
 our theoretical proposal with these experimental outcomes from the ATLAS experiment. \\
In  Fig.~\ref{Fig8}, adapted from Fig.~7 of Ref.~\cite{ATLAS:2022}, we display the observed limits at 95 \% confidence level 
on a heavy Higgs-like  boson production cross section times the branching fraction  $Br(H \rightarrow WW)$ 
 for both the gluon-gluon fusion  and vector-boson fusion production mechanisms in the narrow width approximation.
We also display  our estimate for the product of the gluon-gluon fusion and vector-boson fusion production cross sections  times the branching ratio 
for the decay of  the H Higgs boson into two  vector W bosons.  
\begin{figure}
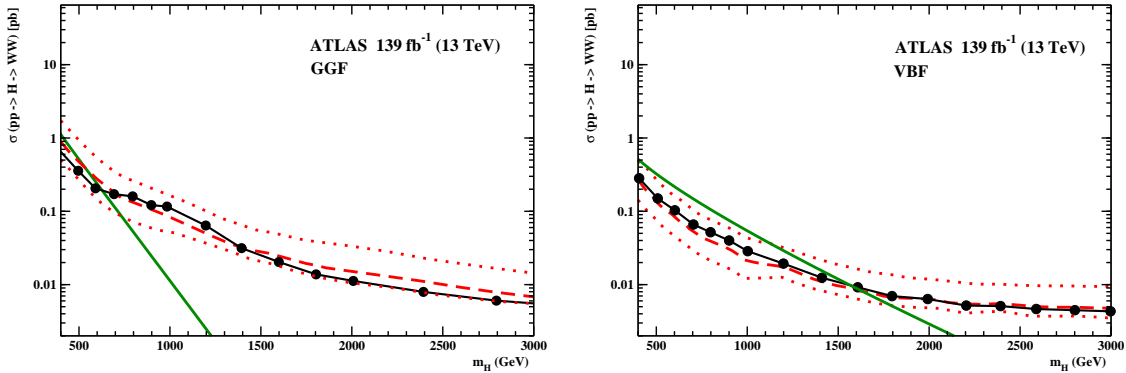

\vspace{-1.0cm}
\begin{center}
\includegraphics[width=0.45\textwidth,clip]{Fig8a.eps}
\hspace{0.2 cm}
\includegraphics[width=0.45\textwidth,clip]{Fig8b.eps}
\end{center}
\caption{\label{Fig8} 
(Color online) Limits on the GGF (left panel) and VBF (right panel) production cross sections times the branching fraction for
the processes $ pp \rightarrow H  \rightarrow  W$.   Full circles represent the observed signal, the (red) dashed lines are
the expected signal, while  the (red) dotted lines demarcate the 95 \% confidence level regions of the expected background. 
The continuous (green) lines are our theoretical estimate for the gluon-gluon fusion  and 
vector-boson fusion production cross sections times the branching ratio Br($H \rightarrow WW$). 
 }
\end{figure}
In the comparison of our theoretical results with the observed limits one should keep in mind that 
the narrow width approximation adopted by  the ATLAS Collaboration  amounts to set the decay width of the heavy scalar
 to be much smaller than the detector resolution and the our branching ration for the decay $H \rightarrow WW$ is slightly higher
 than the one obtained within  the usual perturbative approximation with the classical Mexican-heat Higgs potential.
 As a consequence the experimental limits become somewhat weaker in the case our  heavy Higgs broad resonance. 
Looking at Fig.~\ref{Fig8}  it is evident that,  in the case of the gluon-gluon production mechanism  our theoretical cross section is still  compatible with  the observed  limits in the relevant mass range. As concern the vector-boson production mechanism, we see that our theoretical
estimate of the  production cross sections times the branching ratio Br($H \rightarrow WW$) stays higher
than the observed limits in the high-mass region of interest.  Even taking into account the previous caveats, we see that
at best our theoretical results are compatible with the ATLAS data at the 95 \% confidence level. \\
To conclude the present Section,  we can affirm that the LHC full Run 2 datasets from the ATLAS Collaboration did not display a clear evidence for
an additional heavy Higgs boson. Even thought  we already argued~\cite{Cea:2021} that probably there is not enough sensitivity,
mainly due to the severe selection cuts,  to detect the expected signal especially in the vector-boson fusion production mechanism,
we must admit that  the absence of experimental evidences of an additional heavy Higgs boson from the ATLAS experimental
results looks problematic.  We shall further discuss  this point after the comparison with the data collected from  the CMS experiment.
\subsection{Comparison with the CMS Run 2 datasets}
\label{s-5-2}
The complete Run 2 datasets released by the LHC CMS Collaboration correspond to an integrated luminosity of
about 138 fb$^{-1}$.
In our previous paper~\cite{Cea:2019}  we found some evidence  of a broad scalar resonance  that looks  consistent 
with our Standard Model heavy Higgs boson in the golden channel based on the preliminary Run 2 data collected by
the CMS Collaborations corresponding to an integrated luminosity of $77.4$ fb$^{-1}$.
Nevertheless, observing that the main decay mode of a heavy Higgs boson is the decays into two
 W vector bosons,  the lack  of experimental evidences  in this decay channel posed challenging problems to our
 theoretical proposal.  In fact,  we were well aware that  in absence of convincing  experimental signatures in the  decay
 channel  $H \rightarrow WW$ would  lead us to reject the heavy Higgs boson proposal.  
 Fortunately,  the recent paper Ref.~\cite{CMS:2022}, where  the CMS Collaboration presented a search 
 for a high mass resonance decaying into a pair  of W bosons  using the full data  recorded by 
 CMS during the LHC Run 2 and corresponding to an integrated luminosity of 138 fb$^{-1}$, we found some evidences for
 a broad resonance that resembled a heavy Higgs boson~\cite{Cea:2022}.  \\
The  aim of the present Section is to compare our theoretical proposal for an additional heavy Higgs boson to the
full Run 2 datasets released by the CMS Collaboration in the main decay channels $H \rightarrow WW, ZZ$. \\
Concerning the heavy Higgs decay into two W vector bosons, we shall follow closely our recent paper Ref~\cite{Cea:2022}. 
 The CMS Collaboration presented a search for a high mass resonance decaying into a pair 
of W bosons, using the full data set recorded by CMS during the LHC Run 2
corresponding to an integrated luminosity of 138 fb$^{-1}$.  The search strategy for $H \rightarrow W^+W^-$ was
 based on the final state in which  both   W bosons decay leptonically, resulting in a signature with two isolated, 
oppositely charged, high $p_T$ leptons (electrons or muons) and large missing  transverse  momentum,  
due to the undetected neutrinos. So that,  the bulk of the signal comes  from direct W decays to electrons or muons of opposite 
charge. However, even if not explicitly mentioned in Ref.~\cite{CMS:2022}, the small contributions  proceeding through an 
intermediate $\tau$ lepton are implicitly included. Therefore,  in Ref.~\cite{CMS:2022} it is always included  the $W$ boson 
decays into all three lepton types, such as  the $\ell$ symbol in $W \rightarrow \ell + \nu$  comprises all three leptons  $e, \mu, \tau$.
To increase the signal sensitivity, it were used  a event categorisation optimised for the gluon-gluon fusion  and
vector-boson fusion  production mechanisms. Accordingly,  it was introduced
a parameter $f_{VBF}$ corresponding to the fraction of the VBF production cross section 
 with respect to the total cross section. As a result,  $f_{VBF} = 0$ corresponds to GGF production signal
  while  $f_{VBF} = 1$  considers only the VBF production signal. For a heavy scalar resonance
 with Standard Model-like couplings  $f_{VBF}$  was  set to the expectation using
 the cross sections provided by the LHC Higgs Cross SectionWorking Group~\cite{Hcross13Tev}. \\
The results were presented as  upper limits on the product of the cross section with  the relevant branching ratio on the production of a
 high mass  scalar resonance. The 68 \% and 95 \% confidence level upper limits 
 on   $\sigma(p  p   \rightarrow  H \rightarrow WW \rightarrow 2 \ell 2 \nu)$ 
are displayed in  Fig.~4 of Ref.~\cite{CMS:2022} for four different scenarios, $f_{VBF} = 0$, $f_{VBF} = 1$, floating $f_{VBF}$ and
Standard Model $f_{VBF} $. Interestingly enough, the CMS Collaboration reported an  excess of data over
the Standard Model background expectations for  heavy Higgs boson masses ranging in the interval  600 GeV  -  1000 GeV.
Moreover, the largest localised excess in the data over the background were observed  in the VBF production mechanism
 near 650 GeV  with a local significance of about 3.8 standard deviations. Note, however, that due to the large fraction of missing energy
  from the neutrinos, the resolution on the mass is still quite broad. As a consequence, the excess is
 compatible with a resonant signal with mass ranging from roughly 600 GeV to 1000 GeV. \\
To be concrete, in order to compare with our theoretical expectations
 in Fig.~\ref{Fig9} we report  $\sigma(p  p   \rightarrow  H \rightarrow WW \rightarrow 2 \ell 2 \nu)$ as a function of the Higgs mass $m_H$ 
in the case of a heavy scalar resonance with Standard Model-like couplings (Standard Model $f_{VBF}$).
\begin{figure}
\vspace{-0.0cm}
\begin{center}
\vspace{-1.5cm}
\includegraphics[width=0.78\textwidth,clip]{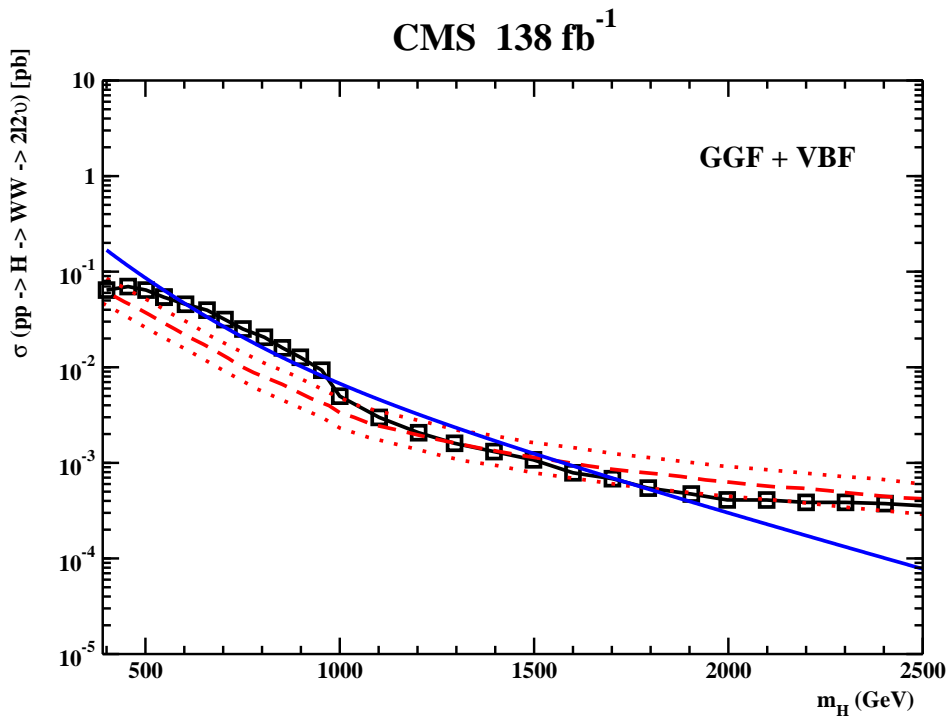}
\end{center}
\caption{\label{Fig9} 
(Color online) Product of the cross section   $\sigma(p  p   \rightarrow  H)$  with  the  branching ratio  $Br(H \rightarrow WW \rightarrow 2 \ell 2 \nu)$
for the search of  a Standard Model heavy Higgs boson  versus the Higgs mass $m_H$. 
The data have been adapted from Fig.~4, bottom right panel, of Ref.~\cite{CMS:2022}.  Black squares correspond to the observed
signal, the (red) dashed line is the expected Standard Model background together with the 68 \% CL limits (red dotted lines).
The full (blue)  line corresponds to the product of the cross section times the branching ratio for our heavy Higgs boson.
}
\end{figure}
Looking at  Fig.~\ref{Fig9} it is evident that  the observed signals display a sizeable broad excess  with respect to expected Standard Model signal in 
the mass range 600 GeV - 800 GeV.  Clearly, these excesses cannot be accounted for by a heavy scalar resonance with a narrow width.
In addition, for a heavy Standard Model Higgs boson the main production mechanism would be  by gluon-fusion processes for 
$m_H \lesssim  1000$ GeV, so that the resulting production cross section  would led to a  signal greater by at least a factor of two with respect
to the observed signal (see red line in Fig.~4, bottom right panel, of Ref.~\cite{CMS:2022}).
 On the other hand, in our theoretical proposal the heavy Standard Model Higgs boson has a rather large width  implying that the expected 
 signal extends on the mass range 600 GeV - 800 GeV with a broad peak around $m_H \simeq 700$ GeV. Moreover, we already argued
 that the main production mechanism is by vector-boson fusion since the gluon-gluon fusion processes are strongly suppressed
 (see Fig.~\ref{Fig6}). To be quantitative, using Eq.~(\ref{5.6}) for the  cross section
  and Eqs.~(\ref{5.8}) - (\ref{5.11}) to evaluate the relevant branching ratio,  we may easily estimate the inclusive production cross
 section for our heavy Higgs boson in the given channels. 
 The result, displayed in Fig.~\ref{Fig9}, seems to compare reasonable well  to the observed signal
 in the relevant Higgs mass range 600 GeV $  \lesssim m_H  \lesssim$  800 GeV.
It should be emphasised that the  rejection of the  background-only hypothesis in a statistical sense  will depend,
 in general, on  the plausibility of the new signal  hypothesis and the degree to which it can describe the data. 
In this respect, the presence of a rather broad excess around  $m_H \simeq 700$ GeV is perfectly consistent
with the fact that our Standard Model heavy Higgs boson has a central mass at   $m_H \simeq 730$ GeV and
a huge width. Moreover, we have estimated  that the cumulative effects of the excesses in the mass range 
600 GeV - 800 GeV reach a statistical significance well above five standard deviations even including the look elsewhere effect~\cite{Gross:2010}.
The look elsewhere effect takes into account that,  when searching for a new resonance 
somewhere in a possible mass range, the significance of observing a local excess of events must take into account 
the probability of observing such an excess anywhere in the range.  Note, however, that for a broad resonance
this effect should not matter due to the fact that statistical fluctuations over an extended mass range are practically
never realised. 
It is necessary to mention that the agreement of our theoretical estimates with the observed production cross section limits depends crucially
on the circumstance that our heavy Higgs boson H is produced mainly by the  vector boson fusion mechanism and that the main
decay modes are the decays into two massive gauge vector bosons. \\
This preliminary evidence of a broad scalar resonance from the CMS observations looks encouraging. However, we have said in the previous 
Section that the full Run 2 datasets from the ATLAS Collaboration data did not displayed experimental evidences of an additional heavy Higgs boson
notwithstanding both experiments used  a comparable  integrated luminosity. To better elucidate this last point, it is useful to directly
compare the  limits on the GGF and VBF production cross sections times the branching fraction for the processes
 $ pp \rightarrow H  \rightarrow  WW$ reported by the two experiments. %
\begin{figure}
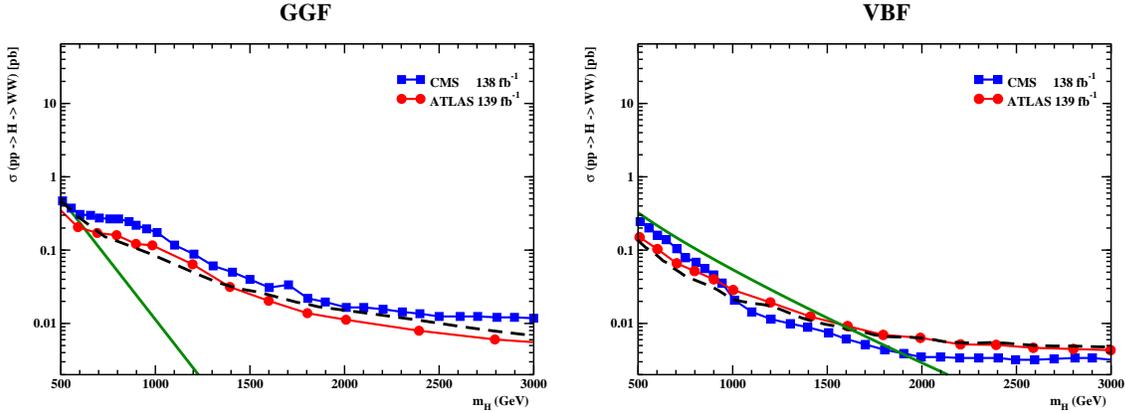

\vspace{-1.0cm}
\begin{center}
\includegraphics[width=0.45\textwidth,clip]{Fig9aBis.eps}
\hspace{0.2 cm}
\includegraphics[width=0.45\textwidth,clip]{Fig9bBis.eps}
\end{center}
\caption{\label{Fig9Bis} 
(Color online) Limits on the GGF (left panel) and VBF (right panel) production cross sections times the branching fraction for
the processes $ pp \rightarrow H  \rightarrow  W$.   Full (red) circles represent the observed signal from ATLAS, 
full (blue) squares  correspond to the observed signal from CMS. The (black) dashed lines are
the expected signal from ATLAS. 
The continuous (green) lines are our theoretical estimate for the gluon-gluon fusion  and 
vector-boson fusion production cross sections times the branching ratio Br($H \rightarrow WW$). 
 }
\end{figure}
\\
In Fig.~\ref{Fig9Bis} we display  the limits on the gluon-gluon fusion  (left panel) and vector-boson fusion(right panel) production cross sections 
times the relevant branching fraction  reported from ATLAS and CMS experiments.
For the  CMS  limits we used: 
\begin{equation}
\label{5.16}
Br(W \, W \;  \rightarrow \; \ell \,  \nu \,  \ell \,  \nu)   \; \simeq  \;  0.103 \; \;, \; \;  \ell \; = \; e, \,  \mu, \,  \tau \; . 
\end{equation}
Figure ~\ref{Fig9Bis}  show clearly that in the mass range 600 GeV - 800 GeV the CMS limits stay above the ones from the ATLAS by
about a factor of two.  In our opinion these discrepancies arise from the circumstance that the tight selection cuts adopted by the
ATLAS Collaboration resulted in a smaller sensitivity for the search of a broad scalar resonance in the mass range relevant to us.  
Obviously, it goes without saying that this matter can be elucidated only  by the experimental collaborations.
\\
We are well aware that the above preliminary experimental evidences
of the H Higgs boson are not enough to claim the  experimental discovery of a new particle.  Fortunately we have at our disposal
the already mentioned   evidence  of a broad scalar resonance  that looks  consistent  with the  heavy Higgs boson in the golden channel
based on the preliminary Run 2 data collected by the CMS collaborations~\cite{Cea:2019}.
\begin{figure}
\vspace{-4.0cm}
\begin{center}
\includegraphics[width=0.8\textwidth,clip]{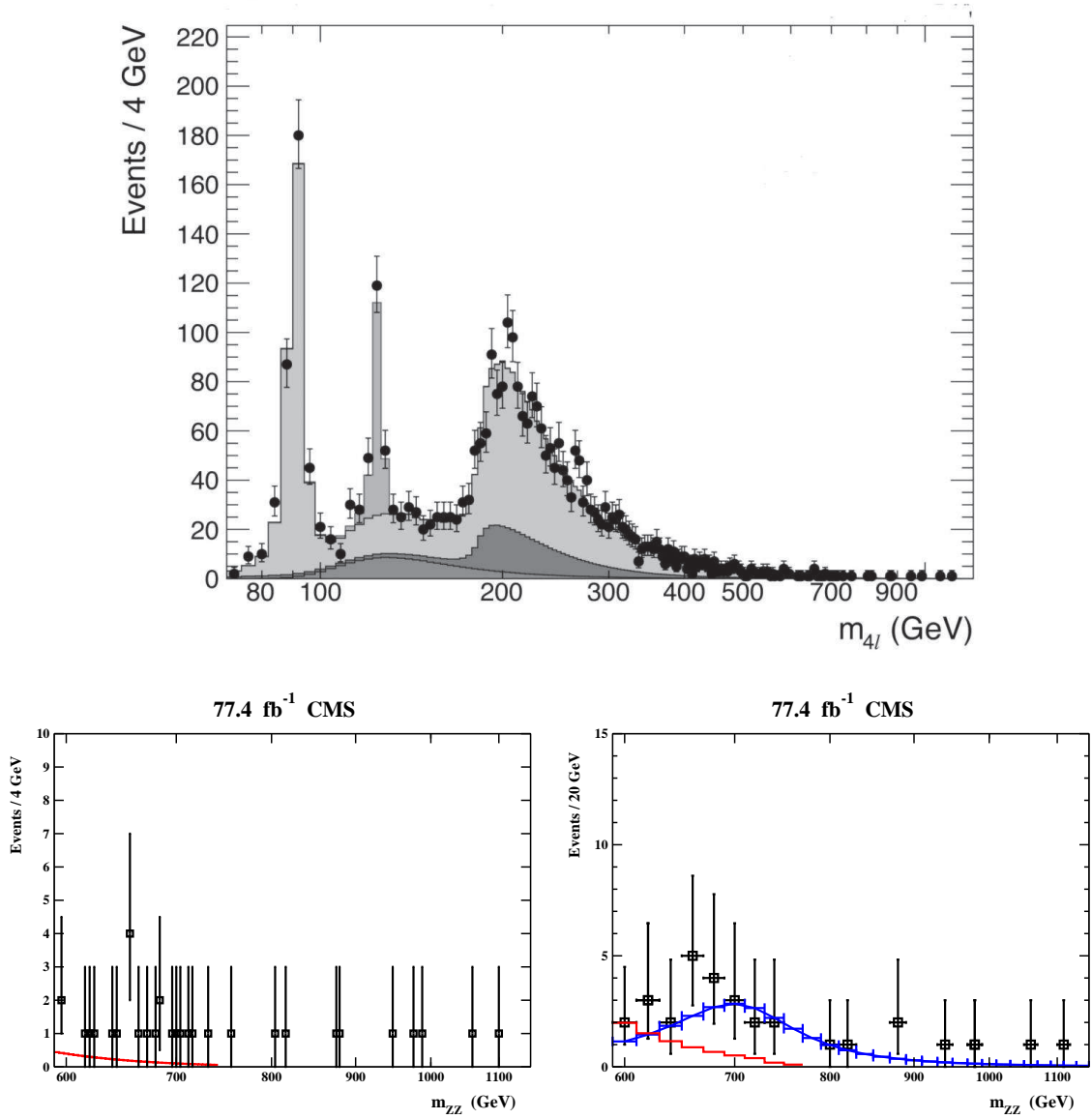}
\\
\vspace{0.2cm}
\includegraphics[width=0.45\textwidth,clip]{Fig11a.eps}
\hspace{0.2 cm}
\includegraphics[width=0.45\textwidth,clip]{Fig11b.eps} 
\end{center}
\caption{\label{Fig11} 
(Color online) 
(Upper panel) Distribution of the four-lepton reconstructed invariant mass $m_{4 \ell}$ in the full mass range
combining the 2016 and 2017 CMS data. Adapted from Fig. 9 of Ref.~\cite{CMS:2018}. 
Distribution of the invariant mass $m_{Z Z}$, with bin size 4 GeV (bottom left  panel) and 20 GeV (bottom right  panel) for the  process
 $H  \; \rightarrow ZZ \; \rightarrow \ell \ell \ell \ell$   ($\ell = e, \mu$)  in the high-mass region   $m_{Z Z}  \gtrsim 600 \,  GeV$.
Adapted from Fig.~1, left panels, of Ref.~\cite{Cea:2019}. The (red) lines are the expected Standard Model background.
The (blue) line is  the expected signal histogram for our heavy Higgs boson.}
\end{figure}
It is useful to summarise the strategy we followed in Ref.~\cite{Cea:2019}. Firstly, in Fig.~\ref{Fig11}, upper panel, we reproduce the 
 distribution of the four-lepton reconstructed invariant mass $m_{4 \ell}$  for the golden channel in the full mass range presented
 by the CMS Collaboration by combining  the 2016 and 2017 CMS data~\cite{CMS:2018} and corresponding to an integrated luminosity
 of 77.4 fb$^{-1}$. From the full range distribution we extracted the  invariant mass distribution   with bin size 4 GeV
  in the high-mass region   $m_{Z Z}  \gtrsim 600 \,  GeV$. This high-mass distribution is displayed in Fig.~\ref{Fig11}, bottom left  panel.
 There is also displayed the expected Standard Model background mainly due to  processes $ gg, q \bar{q} \rightarrow ZZ$.
Observing that the expected Standard Model  background is driven by sea partons, the ZZ  invariant mass distribution should be exponentially 
suppressed in the high-mass region $m_{Z Z}  \gtrsim 500 \,  GeV$. Indeed, this is clearly reproduced by numerical Monte Carlo 
simulations (see the red continuous lines in Fig.~\ref{Fig11}).
From Fig.~\ref{Fig11}, bottom left panel, it should be evident that the number of events above   $m_{Z Z}  \gtrsim 600 \,  GeV$ far exceeds 
 the expected background. In addition,  the events seem to be more densely distributed in the region around 700 GeV. This
 could signal the presence of a resonance with width $\Gamma$ much more higher than the bin size  4 GeV. To enlighten this we have
 binned the events with a bin size of 20 GeV. The result, displayed in Fig.~\ref{Fig11}, bottom right panel, does indeed support the evidence
 of a broad structure around 700 GeV. In  Fig.~\ref{Fig11}  we also compare the theoretical invariant-mass distribution
 of the heavy H Higgs boson to the experimental data. At this point it is necessary to precise that in our theoretical proposal,
once  the masses of the light Higgs boson $h$ and the heavy Higgs boson $H$ are fixed, $m_h \simeq 125$ GeV and $m_H \simeq 730$ GeV,
there are no free parameters left to be tuned to better match the experimental observations. So that, the invariant-mass distribution
in  Fig.~\ref{Fig11}, bottom right panel,  must be considered a prediction of our theory. 
Actually,  we see that the theoretical distribution compares rather well with observations. Nevertheless, as a further check, we must wait for
 the release of the complete Run 2 dataset in the golden channel  where the integrated luminosity  should had increased by about a factor of two. 
\\
Looking at  Figs.~\ref{Fig9} and  \ref{Fig11} one could be led  to the claim  that the CMS Collaboration had found evidences for
 a new heavy resonance with central mass at 730 GeV and a rather large width. However, we do not have at our disposal the relevant LHC
Run 2 datasets. Therefore, our phenomenological analysis, at best, should be considered  as an indication of the presence of a new
heavy resonance.  Actually, the CMS Collaboration can easily  perform full-fledged statistical analyses that can confirm or reject our theoretical proposal.  On the other hand, it should be stressed that the results from the ATLAS Collaboration are not confirming
the presence of a new heavy Higgs-like resonance. We  already suggested that the effects of the
tight   selection cuts applied by the ATLAS Collaboration  on the full Run 2 data could drastically reduce that sensitivity
in the search of a broad heavy scalar resonance in the mass range 500 GeV - 1000 GeV.  As a consequence we must wait for
more data to settle the question. \\ 
We would like to conclude the present Section by  emphasising  that our heavy H resonance resembles  a Standard 
Model Higgs boson since the inclusive production mechanisms are mainly by vector-boson fusion and gluon-gluon fusion
processes. In addition, the main decay modes are the decays $H \rightarrow WW, ZZ$ with branching ratios satisfying 
Eq.~(\ref{5.12}).
\section{Summary and conclusions}
\label{s-6}
The Standard Model of the  Elementary Particle Physics has been confirmed to a high level of precision by the experimental outcomes from
the Large Hadron Collider. In particular, the discovery of a new narrow scalar resonance with mass 125 GeV, that resembles closely
the Higgs boson, had firmly confirmed the Higgs mechanism, namely the spontaneous symmetry breaking of the electroweak
gauge group by scalar fields. Nevertheless, from the theoretical point of view the widely accepted implementation of the Higgs
mechanism is highly unsatisfying. In the perturbative approach to the spontaneous symmetry breaking of the gauge group one
needs scalar fields with tachyonic mass term and positive quartic self-coupling. We repeatedly   stressed  that, from one hand there
are no known physical mechanisms able to generate a negative mass squared for quantum scalar fields, on the other hand
the triviality of self-interacting scalar quantum  field theories invalidates the semi-classical perturbative implementation of the
symmetry breaking. The unique way out left would be to consider the quantum scalar fields non-elementary or compound.
Nevertheless, the absence of experimental evidences for physics beyond the Standard Model led us to esclude these possibilities. \\
In the first part of the paper we have illustrated the triviality and spontaneous symmetry breaking picture for the simplest
scalar quantum field theory, namely the one-component scalar field. We showed that in  that approach the spontaneous symmetry breaking
could be driven by non-Gaussian quantum fluctuation despite the absence of a positive quartic self-coupling. 
However, we also found that that triviality and spontaneous symmetry breaking scenario cannot be realised   within the pure
scalar sector of the Standard Model. After that, we have carefully compared the perturbative  and  triviality and spontaneous symmetry breaking 
approaches to precise and extensive numerical results in the broken phase from the non-perturbative approach offered by the lattice formulation
of the one-component scalar quantum field theory. We have shown that the proposal of triviality and spontaneous symmetry breaking 
compared rather well   with numerical simulations, while the renormalised perturbative approach displayed  statistically significant deviations
with respect to the numerical results. \\
In the second part of the paper we addressed the problem of the spontaneous symmetry breaking in the Standard Model.
As it is well known, in the Standard Model the presence of local gauge symmetries
 implies that to have the spontaneous symmetry breaking  one is forced to fix the 
 gauge. After that, in the unitary gauge the dynamics of the Standard Model scalar sector 
 reduces to the one  of the one-component scalar quantum field. Moreover, we found that 
 the couplings of the  scalar fields to the gauge vector bosons and fermions induce the 
 needed non-Gaussian quantum fluctuations that, in turns, lead to the Bose-Einstein 
 condensation. As a consequence,  we obtained  a clear physical picture of the
 Higgs mechanisms that is not subjected to the stability problem that affects
 the quantum  vacuum in the perturbative approach. Moreover, we argued the 
 the scalar Higgs condensate behave like a relativistic quantum
 liquid leading to the prevision of two Higgs bosons. The light Higgs boson resembles
 closely the new LHC narrow resonance at 125 GeV, while  the heavy Higgs boson
 turns out to be  a rather broad resonance with central mass of about 730 GeV.
 After discussing the  phenomenological signatures of the heavy Higgs boson,
 we looked at eventual  experimental evidences by critically comparing
 our proposal to the complete LHC Run 2 data collected by the ATLAS and CMS
 Collaborations. We did not  found convincingly evidences of the heavy Higgs boson 
 in the ATLAS data.  The CMS data,  on the contrary,   displayed evidences of a heavy Higgs 
 boson in the main decay channels $H \rightarrow WW, ZZ$   with  reasonable
  statistical significance that  compared favourably to our theoretical proposal.
  We, also,  suggested that the discrepancies between the two experiments could
 originate from the severe selection cuts applied by the ATLAS Collaboration in they
 search of a heavy scalar resonance resembling a heavy Higgs boson. \\
In conclusions,  the results presented in the present paper  should have laid a solid theoretical basis for the eventual presence of an additional
heavy Higgs boson.  Of course, our theoretical proposal  must wait for further  LHC data to be validated. Nevertheless,
we hope that our  results  will induce  the LHC ATLAS and CMS  Collaborations  to  undertake 
 carefully searches of an eventual additional massive Higgs boson.   

\end{document}